\documentclass[11pt,preprint]{aastex}

\begin{document}

\newcommand{\mnhi}{N_{\rm HI}}
\newcommand{\nhi}{$N_{\rm HI}$}
\newcommand{\secdustdepl}{3}
\newcommand{\secdustobsc}{4}
\newcommand{\secZn}{2.1.6}
\newcommand{\ZSi}{<Z>^{\rm Si}}
\newcommand{\ZZn}{<Z>^{\rm Zn}}
\newcommand{\ZFe}{<Z>^{\rm Fe}}
\newcommand{\ew}{W_\lambda}
\newcommand{\XFe}[1]{<[{\rm #1/Fe}]>}
\newcommand{\feh}{[{\rm Fe/H}]}
\newcommand{\znh}{[{\rm Zn/H}]}
\newcommand{\ndla}{37}
\newcommand{\msol}{M_\odot}
\newcommand{\pkts}{P_{KS}}
\newcommand{\etal}{et al.\ }
\newcommand{\delv}{\Delta v}
\newcommand{\Lya}{Ly$\alpha$}
\newcommand{\lya}{Ly$\alpha$}
\newcommand{\lyb}{Ly$\beta$}
\newcommand{\lyg}{Ly$\gamma$}
\newcommand{\lyd}{Ly$\delta$}
\newcommand{\lye}{Ly$\epsilon$}
\newcommand{\ndmp}{31 }
\newcommand{\nnew}{14 }
\newcommand{\nrun}{8,500}
\newcommand{\kms}{km~s$^{-1}$ }
\newcommand{\cm}[1]{\, {\rm cm^{#1}}}
\newcommand{\N}[1]{{N({\rm #1})}}
\newcommand{\f}[1]{{f_{\rm #1}}}
\newcommand{\e}[1]{{\epsilon({\rm #1})}}
\newcommand{\rAA}{{\AA \enskip}}
\newcommand{\sci}[1]{{\rm \; \times \; 10^{#1}}}
\newcommand{\ltk}{\left [ \,}
\newcommand{\ltp}{\left ( \,}
\newcommand{\ltb}{\left \{ \,}
\newcommand{\rtk}{\, \right  ] }
\newcommand{\rtp}{\, \right  ) }
\newcommand{\rtb}{\, \right \} }
\newcommand{\ohf}{{1 \over 2}}
\newcommand{\nohf}{{-1 \over 2}}
\newcommand{\rhf}{{3 \over 2}}
\newcommand{\smm}{\sum\limits}
\newcommand{\perd}{\;\;\; .}
\newcommand{\cmma}{\;\;\; ,}
\newcommand{\sgint}{\sigma_{int}}
\newcommand{\Nperp}{N_{\perp} (0)}
\newcommand{\intl}{\int\limits}
\newcommand{\und}[1]{{\rm \underline{#1}}}
\newcommand{\sphr}{\sqrt{R^2 + Z^2}}
\newcommand{\vrot}{v_{rot}}
\newcommand{\btau}{\bar\tau (v_{pk}) / \sigma (\bar \tau)}
\newcommand{\Ipk}{I(v_{pk})/I_{c}}
\newcommand{\mkms}{{\rm \; km\;s^{-1}}}
\newcommand{\ovi}{{\rm O~VI}}
\newcommand{\EW}{W_\lambda}

\def\lick{1}
\def\uc{2}
\def\berk{3}
\def\geneve{4}
\def\scranton{5}
\def\michigan{6}
\def\mit{7}
\def\harvard{8}
\def\magellan{9}
\def\ociw{10}

\title{The Interstellar Medium of GRB Host Galaxies I.
Echelle Spectra of Swift GRB Afterglows
}

\author{ J.X. Prochaska\altaffilmark{\lick},
        H.-W. Chen\altaffilmark{\uc}, 
        J. S. Bloom\altaffilmark{\berk},
	M. Dessauges-Zavadsky\altaffilmark{\geneve}, 
	J. M. O'Meara\altaffilmark{\scranton}, 
	R. J. Foley\altaffilmark{\berk}, 
	R. Bernstein\altaffilmark{\michigan}, 
	S. Burles\altaffilmark{\mit}, 
        A. K. Dupree\altaffilmark{\harvard,\magellan}, 
        E. Falco\altaffilmark{\harvard}, 
	I. B. Thompson\altaffilmark{\ociw,\magellan}
        }

\altaffiltext{\lick}{UCO/Lick Observatory; University of California, Santa Cruz;
	Santa Cruz, CA 95064; xavier@ucolick.org}
\altaffiltext{\uc}{Department of Astronomy; University of Chicago;
	5640 S. Ellis Ave., Chicago, IL 60637; hchen@oddjob.uchicago.edu}
 \altaffiltext{\berk}{Department of Astronomy, 601 Campbell Hall, 
        University of California, Berkeley, CA 94720-3411}
 \altaffiltext{\geneve}{Observatoire de Gen\`eve, 51 Ch. des Maillettes, 
	1290 Sauverny, Switzerland}
\altaffiltext{\scranton}{Department of Physics, Penn State 
Worthington Scranton, 120 Ridge View Drive, Dunmore PA 18512
}
\altaffiltext{\michigan}{Department of Astronomy, University of Michigan, 
	Ann Arbor, MI 48109}
\altaffiltext{\mit}{MIT Kavli Institute for Astrophysics and Space Research,
	Massachusetts Institute of Technology, 77 Massachusetts Avenue, 
	Cambridge MA 02139}
 \altaffiltext{\harvard}{Harvard-Smithsonian Center for Astrophysics, 
        60 Garden St., Cambridge, MA 02138}
\altaffiltext{\magellan}{This paper includes data gathered with the 
6.5 meter Magellan Telescopes located at Las Campanas Observatory.}
\altaffiltext{\ociw}{Observatories of the Carnegie Institution of
Washington, 213 Santa Barbara St., Pasadena, CA 91101}

\begin{abstract} 

We present optical echelle spectra of four gamma-ray
burst (GRB) afterglows (GRB~050730, GRB~050820, GRB~051111, and GRB~060418)
discovered during the first 1.5~years of operation of the {\it Swift}
satellite and localized by either
the {\it Swift} telescope or follow-up ground-based imaging.
We analyze the
spectra to derive accurate column density measurements for the
transitions arising in the interstellar medium (ISM) of the GRB host
galaxies.  These measurements can be used to constrain the physical
properties of the ISM including the metallicity, dust-to-gas ratio,
ionization state, and chemical abundances of the gas.
We also present measurements of the strong \ion{Mg}{2} systems
in the GRB afterglow spectra.
With the publication of this paper, we provide the first data
release of echelle afterglow spectra by the GRAASP collaboration 
to the general community.


\end{abstract}

\section{Introduction}

The launch of the {\it Swift} satellite \citep{gcg+04} has 
revolutionized the study of the interstellar medium of gamma-ray
burst (GRB) host
galaxies and fostered new studies on the
intervening intergalactic medium \citep{cpb+05,ppc+06}. 
While previous missions alerted the community to the existence
of very bright GRB afterglows \citep[e.g.][]{kdo+99}, 
only a few events were identified in time to allow 
echelle observations \citep{fdl+05}.
The rapid localization by {\it Swift} of the hard X-ray emission to several
arcminutes \citep{bat05}
and the soft X-ray component to a few arcseconds \citep{xrt05}
has enabled observers to obtain high signal-to-noise,
high-resolution spectroscopy of GRBs with 10m-class telescopes.

In this paper, we present a uniform dataset of echelle spectra
of four GRB afterglows by our GRB Afterglows as Probes 
(GRAASP\footnote{http://www.graasp.org}) collaboration. 
Among other applications, the data presented here can be used to
derive constraints on the physical conditions of the ISM of GRB host galaxies:
the metallicity, dust-to-gas ratio, chemical abundance patterns,
ionization state, the distance of the GRB to the absorbing gas, etc.
Future papers in this series will examine these properties in greater
detail and provide comparison with analogous observations along 
quasar sightlines.
In addition to the data presented here, we have collected several low-resolution
spectra of these and other GRBs and will present that data in a companion
paper \citep{pcb+06} along with ISM abundance measurements obtained by different
analysis techniques more suitable to lower resolution data \citep{pro_cog}.
We also present measurements of the strong \ion{Mg}{2} systems
in the GRB afterglow spectra.
In $\S$~\ref{sec:obs} we summarize the observations and data
reduction procedures.  The methodology is briefly described in 
$\S$~\ref{sec:colm} and the
line-profiles and column density
values are given in $\S$~\ref{sec:grb} and $\S$~\ref{sec:mgii}.

\section{Observations and Data Reduction}
\label{sec:obs}

The GRB afterglow spectra presented here were acquired with either
the High Resolution Echelle Spectrometer \citep[HIRES;][]{vogt94}
on the Keck\,I 10m telescope or the 
Magellan Inamori Kycoera Echelle spectrometer \citep[MIKE;][]{bernstein03} on the
Magellan 6.5m Clay telescope at Las Campanas Observatory.
Table~\ref{tab:obslog} presents a summary of the GRB events, the
afterglow observations,
instrumental configurations, and data quality of the spectra
obtained for the four afterglows for which we have obtained high-resolution
spectra: GRB~050730, GRB~050820, GRB~051111, and GRB~060418.

The data were reduced with the HIRES 
Redux\footnote{http://www2.keck.hawaii.edu/inst/hires/} and 
MIKE Redux\footnote{http://web.mit.edu/$\sim$burles/www/MIKE/}
pipelines \citep{bpb06}.
Briefly, the 2D images were bias subtracted, flat-fielded, 
and wavelength calibrated using standard ThAr and quartz lamp
calibration images.  The data were optimally extracted using
a non-parametric spatial profile and coadded after weighting
by the global SNR$^2$ of each echelle order.  A final, normalized 1D
spectrum was produced by fitting high-order Legendre polynomials
to each echelle order (HIRES) or the fluxed 1D spectrum (MIKE).

We provide additional details related to the specific observations
in the following section.

\section{Methods: Column Density Measurements}
\label{sec:colm}

Following our standard practice for the analysis of interstellar
lines observed in the damped \lya\ systems 
\citep[quasar absorption systems 
with $\mnhi > 2\sci{20}\cm{-2}$][]{pro01,pro03},
we have measured column densities using the apparent optical
depth method \citep[AODM;][]{savage91}.  This technique leads to
non-parametric, accurate results provided the profiles are
resolved.  In cases where one measures multiple transitions for
a single ion (e.g.\ \ion{Ni}{2}~1741, 1751), one
can test for line saturation by examining the column density
values derived as a function of the line-strength.  If there is
no trend of increasing column density with decreasing line-strength,
then the profiles are free of significant `hidden' saturation
\citep{jenkins96}.  We find no indication of hidden saturation
in the GRB studied here.

We report column density values for all transitions detected at
greater than $3\sigma$ statistical significance and report
$1\sigma$ statistical uncertainty in the sections that follow.  
For ions or atoms where multiple transitions are measured, we
report the weighted mean column density and combined statistical
error in that mean.  While the data may formally
yield errors of less than $0.01$\,dex, we recommend that one adopt
a minimum statistical uncertainty of 5\% the column density.
Saturated line-profiles, i.e., those where
the minimum normalized flux of the line-profile
is less than the $1\sigma$ statistical
error, are reported as lower limits.  Finally, undetected or blended
transitions are reported as $3\sigma$ upper limits assuming statistical
uncertainty alone.  Accounting for continuum error, these limits are
more likely $2\sigma$ values.  
We note that uncertainties in weak lines are likely to be dominated
by systematics associated with the continuum placement, and 
therefore adopt a minimum
uncertainty of $5\%$ in our own analysis.

The AODM technique cannot be applied to transitions which are blended with
coincident features.  In cases of particular importance, 
we have derived line-profile fits to the data using the software
package VPFIT, kindly provided by R. Carswell.
Finally, when the \lya\ profile is covered by the spectroscopic
observations, we have fitted this line using 
the {\it x\_fitdla} algorithm within the XIDL 
package\footnote{http://www.ucolick.org/$\sim$xavier/IDL}
\citep[e.g.][]{phw05}.  

The following sections present velocity profiles of 
transitions observed along the GRB sightlines.
The transitions are separated into two groups: those arising
from the ground state (termed resonance lines) and those
arising from excited levels (ambiguously termed fine-structure lines).
In all of the figures, line-blends are indicated by dotted lines.
In cases where the line-blending is extremely severe, we have not
shown the velocity profile.  

The atomic data used here has been taken from recent laboratory
measurements and theoretical calculations when lab measurements are
not available: 
\citep[see][]{morton91,bergeson93,bml94,bergeson96,tripp96,raassen98,morton03}.

\section{The ISM of GRB Host Galaxies}
\label{sec:grb}

\subsection{GRB~050730}

The {\it Swift} BAT instrument
triggered on GRB~050730 at UT 2005 July 30 19:58:23
and shortly after \cite{gcn3704} reported the detection of a bright
optical afterglow ($V = 17.6$ at 199s after the trigger)
at RA=$14^h 08^m 16^s.2$, DEC=$-03^\circ 45' 41''$. 
After waiting for the Sun to set at Las Campanas Observatory,
we began acquiring data with the Magellan telescope under good conditions.
\cite{cpb+05} presented a preliminary analysis of this dataset;
we present the full set of transitions observed.

\begin{figure}
\epsscale{0.85}
\plotone{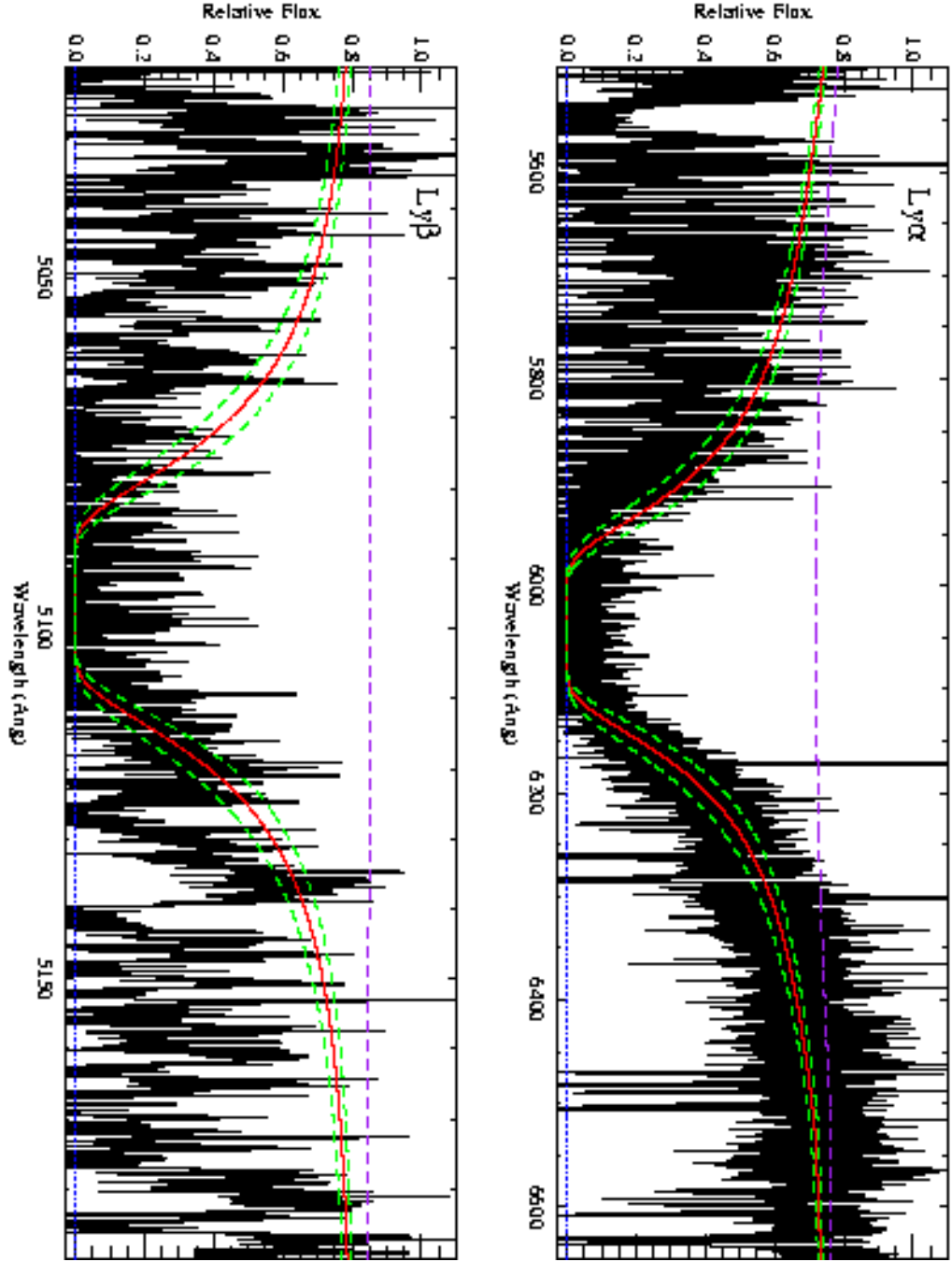}
\caption{\lya\ and \lyb\ transitions for the host galaxy of
GRB~050730.  The dashed purple line above the data shows our
best guess at the afterglow continuum.  The curves overplotted
on the data show the Voigt profiles for our favored solution
($\log \mnhi  = 22.15 \pm 0.10$\,dex).
}
\label{fig:050730lya}
\end{figure}

Figure~\ref{fig:050730lya} presents the \lya\ and \lyb\
profiles for the host galaxy, and the best Voigt profile fit
which gives $\log \mnhi = 21.15 \pm 0.10$dex.  The 
uncertainty in this fit is dominated by the fluxing error associated
with echelle spectrometers and 
continuum fitting.  A lower resolution, well calibrated,
single-order spectrum would likely give a more precise result.

\begin{figure}
\epsscale{0.85}
\plotone{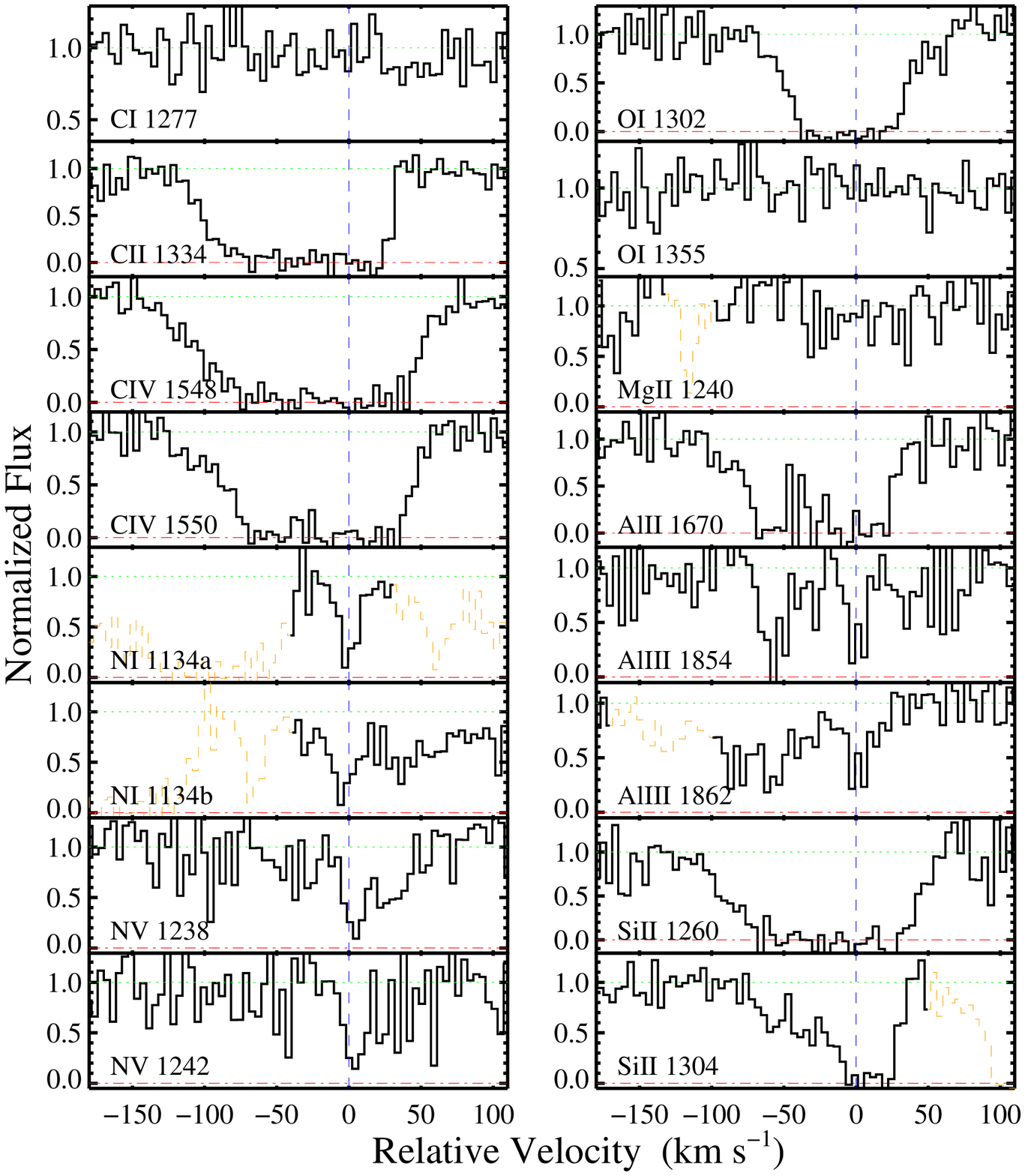}
\caption{Velocity profiles of the resonance transitions identified in the 
ISM of the host galaxy of GRB~050730.  The velocity $v=0\mkms$
corresponds to $z = 3.96855$.  The profiles are ordered by atomic
number and ionization state.
}
\label{fig:050730res}
\end{figure}

\begin{figure}
\epsscale{0.85}
\plotone{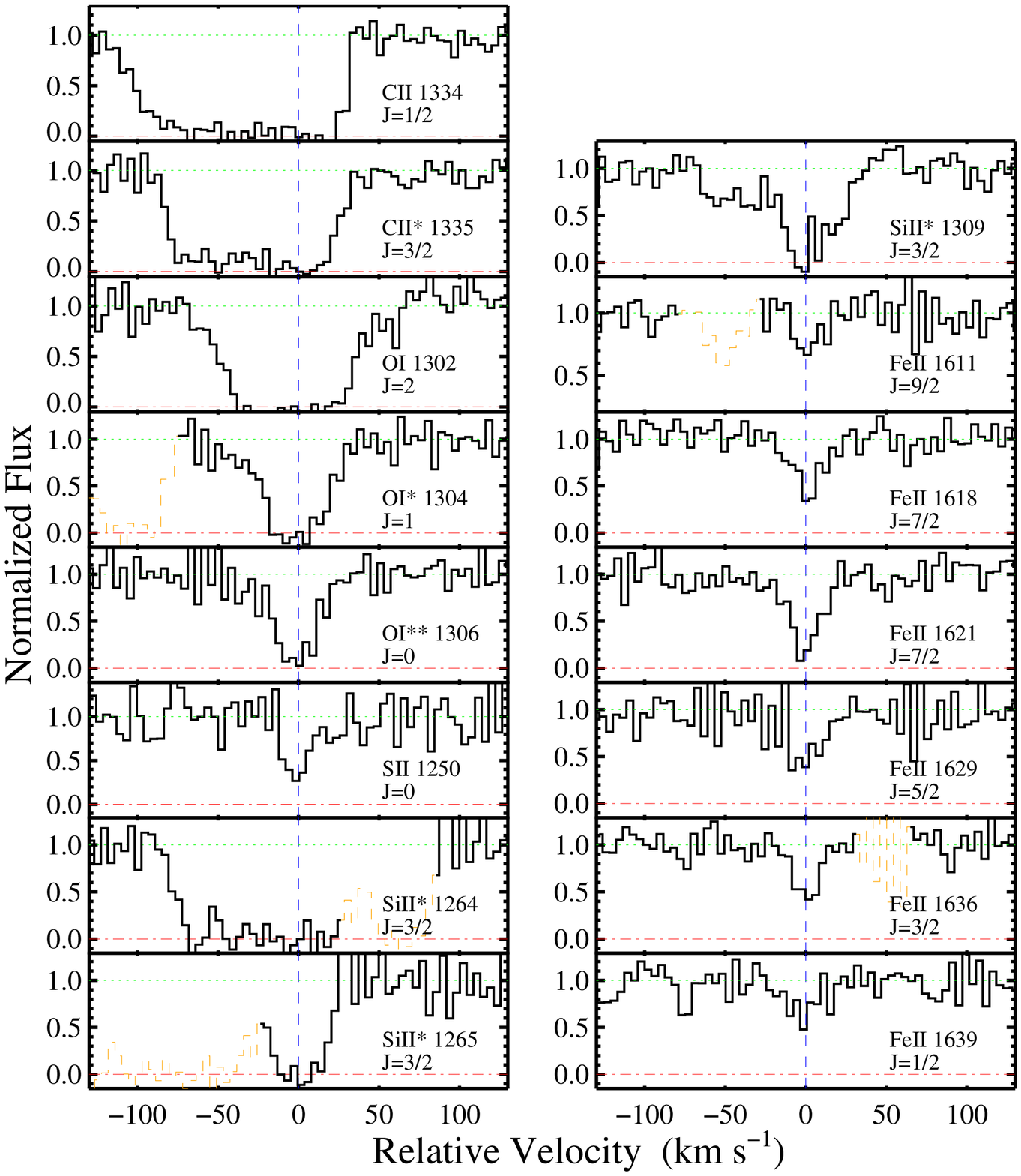}
\caption{Velocity profiles of the fine-structure transitions 
(with several resonance lines for comparison) identified
in the ISM of the host galaxy of GRB~050730.  The velocity $v=0\mkms$
corresponds to $z = 3.96855$.  The profiles are ordered by atomic
number and energy level.
}
\label{fig:050730fine}
\end{figure}

Figure~\ref{fig:050730res} shows the velocity profiles for
the resonance lines associated with the ISM of the host galaxy for
GRB~050730.
Figure~\ref{fig:050730fine} shows the fine-structure transitions
and a resonance line for comparison.  Finally, Table~\ref{tab:050730}
lists the column density measurements for all of the
transitions.

\subsection{GRB~050820}

The {\it Swift} BAT instrument triggered on GRB~050820 at UT 2005 August 20 06:34:53
and the {\it Swift} team
issued an XRT position within a few minutes \citep{gcn3830}.
\cite{gcn3829} reported a bright ($R\sim 14.7$), optical afterglow at 
RA=$22^h 29^m 38^s.11$, DEC=$+19^\circ 33' 37''.1$.  
We began
acquiring data with the HIRES echelle spectrometer on the Keck\,I telescope
shortly after.
Our instrumental configuration followed the standard setup for the UC
planet hunters \citep[e.g.][]{vbm+05}.  The short slit length ($3''$) 
complicated sky subtraction although the seeing ($0.7''$)
was sufficiently good to give fair results.

\begin{figure}
\epsscale{0.85}
\plotone{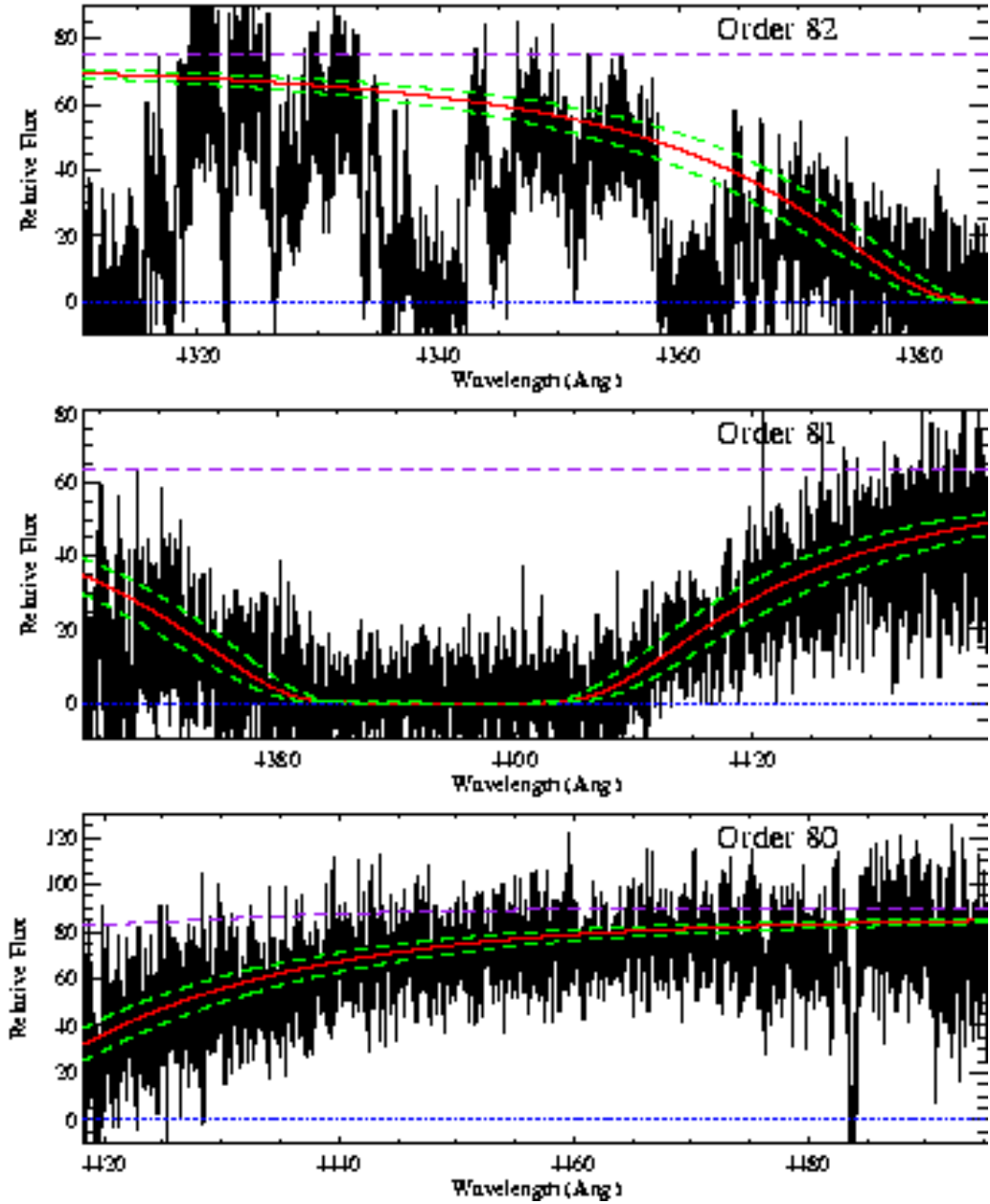}
\caption{Echelle orders covering the \lya\ \ transition for the host galaxy of
GRB~050820.  The dashed purple line above the data shows our
best guess at the afterglow continuum.  The curves overplotted
on the data show the Voigt profiles for our favored solution
($\log \mnhi  = 21.00 \pm 0.10$\,dex).
}
\label{fig:050820lya}
\end{figure}

In Figure~\ref{fig:050820lya}, we present the echelle orders
covering the \lya\ profile for the host galaxy of GRB~050820.  
Overplotted is a fit to the data 
which yields $\log \mnhi = 21.0 \pm 0.1$dex.
The uncertainty in this fit is dominated by continuum error
which is related to difficulties in fluxing this echelle data.

\begin{figure}
\epsscale{0.85}
\plotone{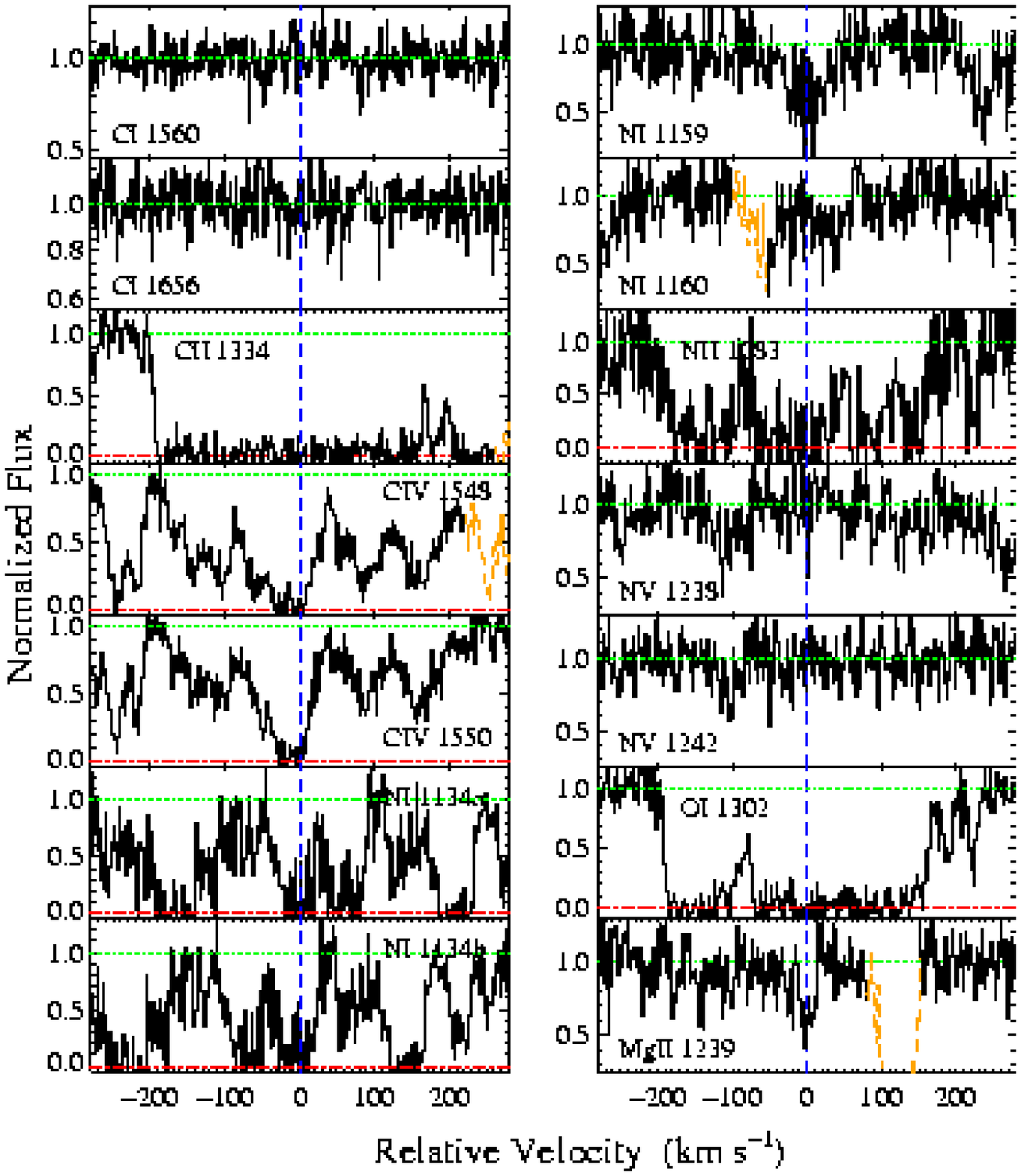}
\caption{Velocity profiles of the resonance transitions identified in the 
ISM of the host galaxy of GRB~050820.  The velocity $v=0\mkms$
corresponds to $z = 2.61469$.  The profiles are ordered by atomic
number and ionization state.
}
\label{fig:050820res}
\end{figure}

\begin{figure}
\epsscale{0.85}
\plotone{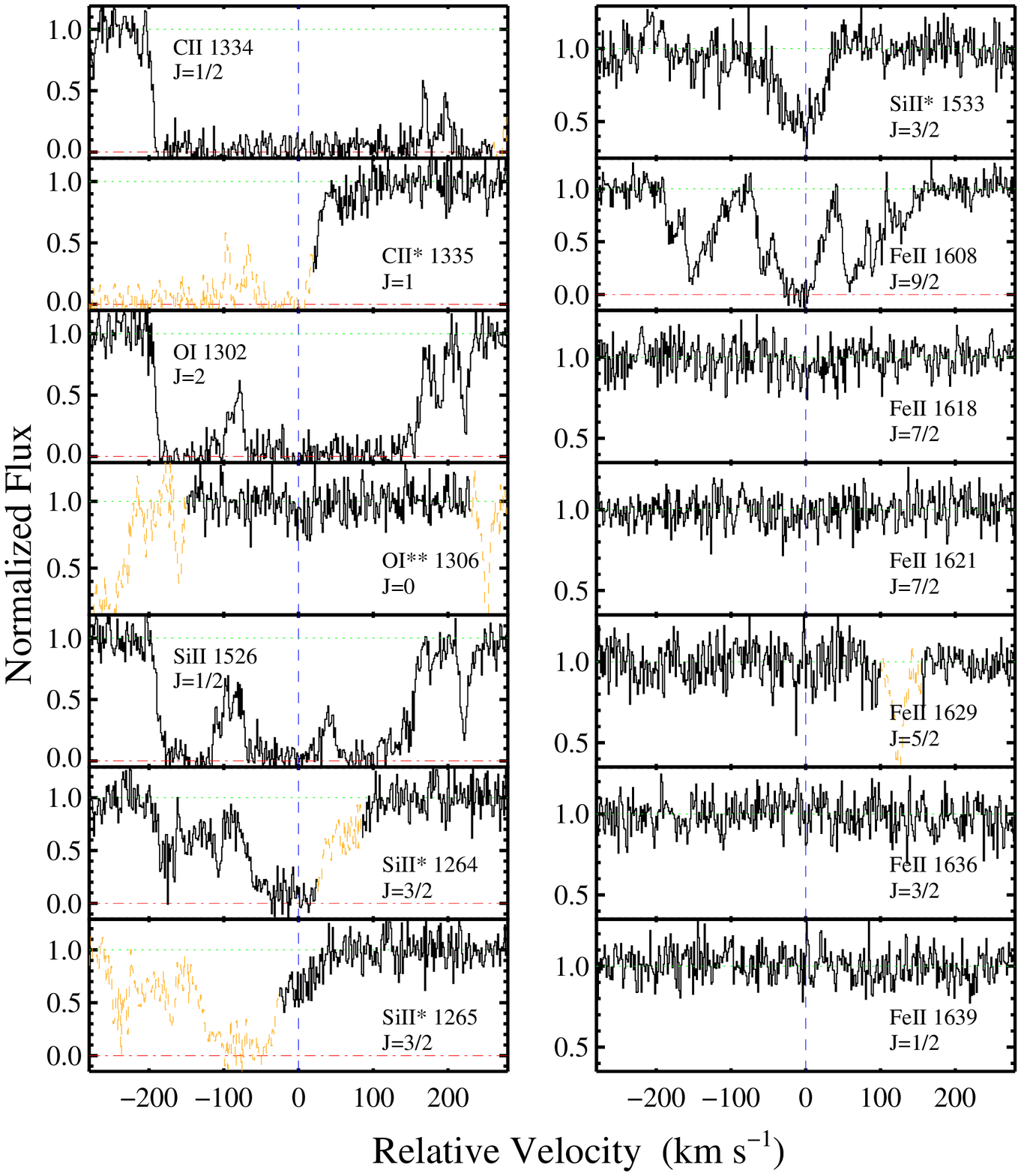}
\caption{Velocity profiles of the fine-structure transitions 
(with several resonance lines for comparison) identified
in the ISM of the host galaxy of GRB~050820.  The velocity $v=0\mkms$
corresponds to $z = 2.61469$.  The profiles are ordered by atomic
number and energy level.
}
\label{fig:050820fine}
\end{figure}

Figure~\ref{fig:050820res} shows the velocity profiles for
the resonance lines associated with the ISM of the host galaxy.
Figure~\ref{fig:050820fine} shows the fine-structure transitions
and a resonance line for comparison.  Finally, Table~\ref{tab:050820}
lists the column density measurements for all of the
transitions.

\subsection{GRB~051111}

The {\it Swift} BAT instrument
triggered on GRB~051111 at UT 2005 November 11 05:59:39 \citep{gcn4299}.  
Owing to the moon constraint, XRT observations were delayed until 5000s
after the trigger \citep{gcn4261}.
\cite{gcn4247} reported an optical afterglow 
at RA=$23^h 12^m 33^s.2$, DEC=$+18^\circ 22' 29''.1$.
Although Keck~I was scheduled for engineering on this night, the observatory
kindly agreed to switch instruments to HIRES and obtain echelle
spectroscopy of GRB~050820 \citep{gcn4255}.

\begin{figure}
\epsscale{0.85}
\plotone{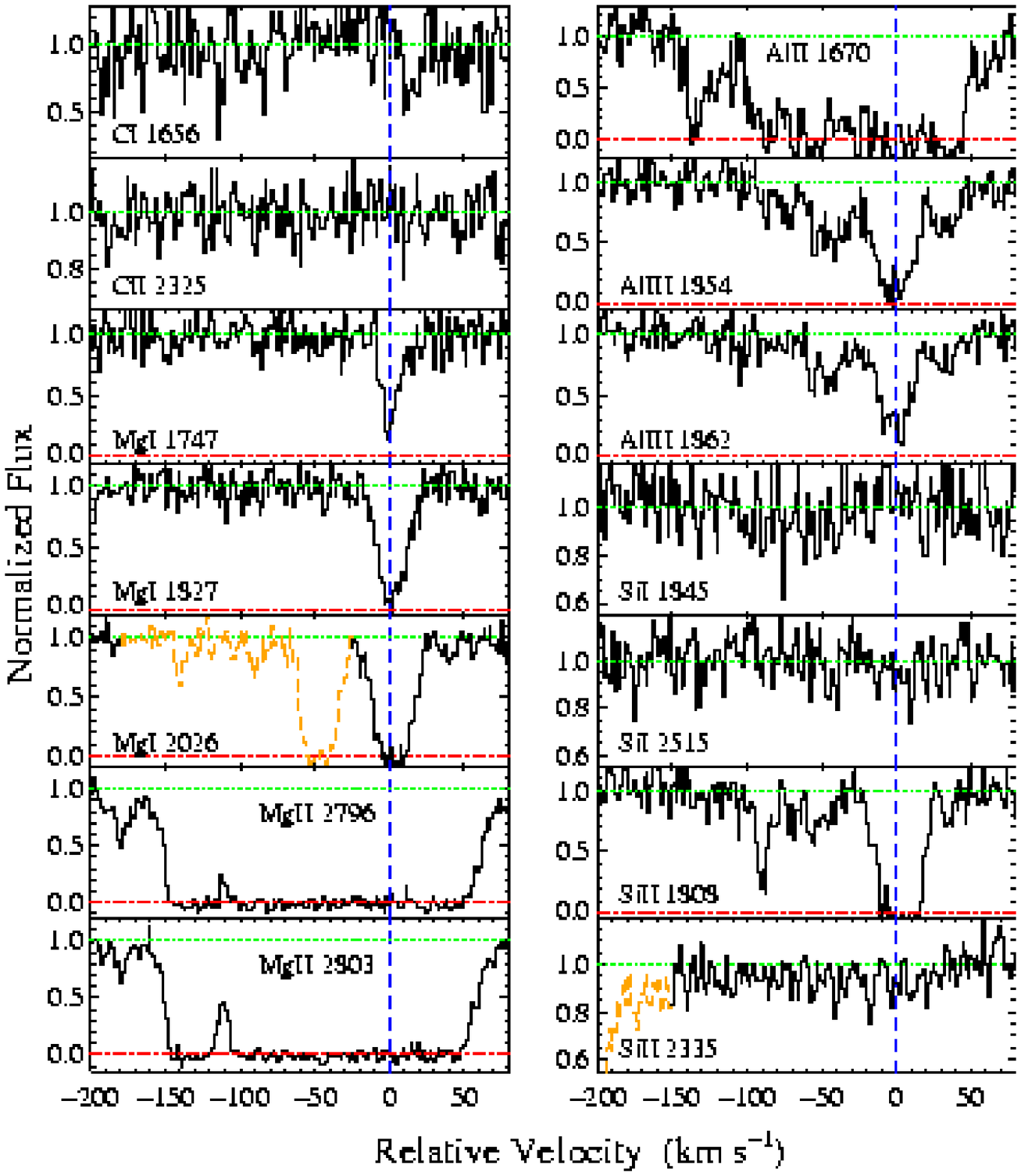}
\caption{Velocity profiles of the resonance transitions identified in the 
ISM of the host galaxy of GRB~051111.  The velocity $v=0\mkms$
corresponds to $z = 1.54948$.  The profiles are ordered by atomic
number and ionization state.
}
\label{fig:051111res}
\end{figure}

\begin{figure}
\epsscale{0.85}
\plotone{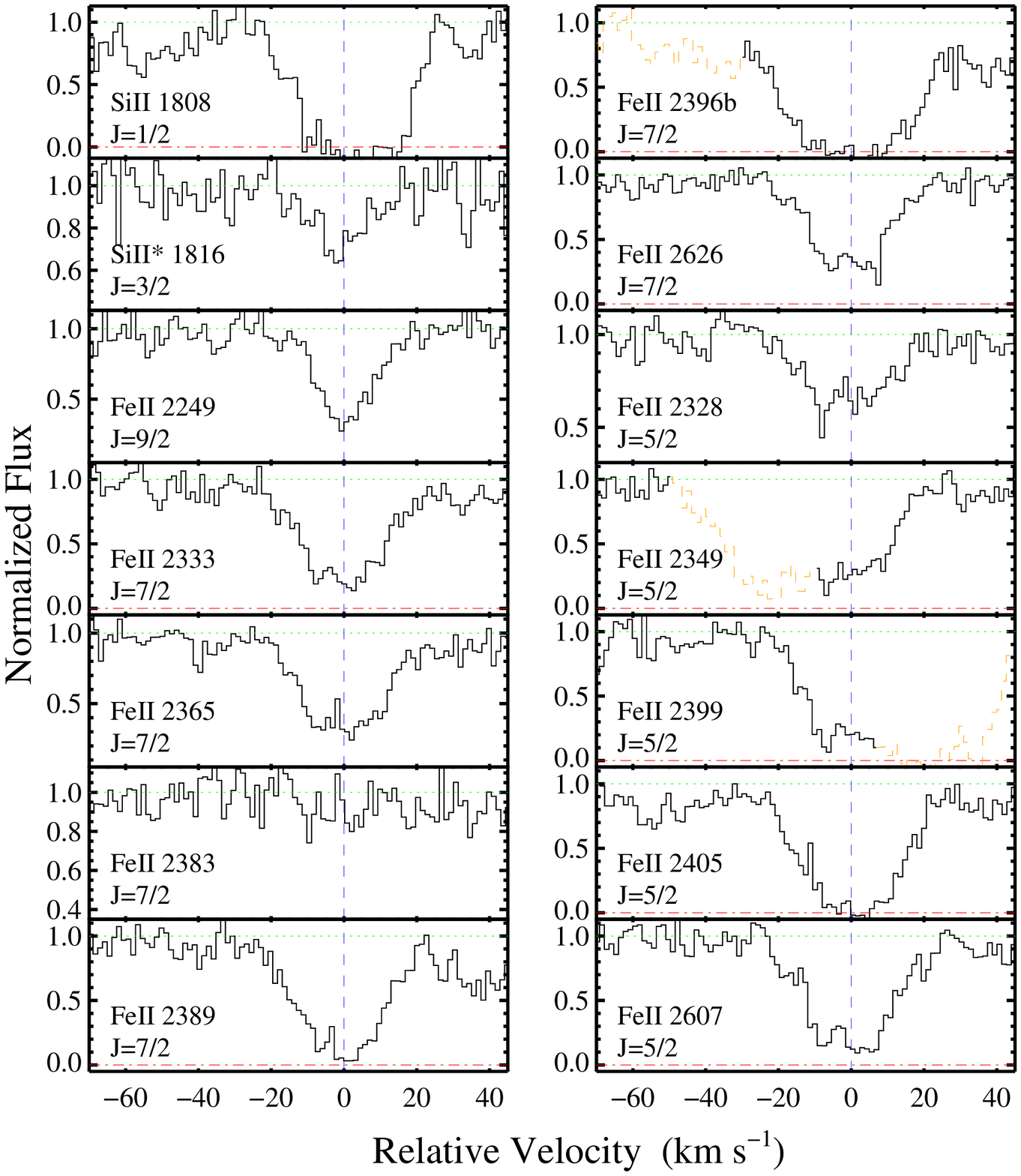}
\caption{Velocity profiles of the fine-structure transitions 
(with several resonance lines for comparison) identified
in the ISM of the host galaxy of GRB~051111.  The velocity $v=0\mkms$
corresponds to $z = 1.54948$.  The profiles are ordered by atomic
number and energy level.
}
\label{fig:051111fine}
\end{figure}

Figure~\ref{fig:051111res} shows the velocity profiles for
the resonance lines associated with the ISM of the host galaxy.
Figure~\ref{fig:051111fine} shows the fine-structure transitions
and a resonance line for comparison.  Finally, Table~\ref{tab:051111}
lists the column density measurements for all of the
transitions.
\cite{penprase05} have also presented an analysis of the same HIRES data 
acquired for this sightline.
Although our central values are generally in good agreement with their
results, we report significantly smaller uncertainties, perhaps
because of differences in coadding the data.

\subsection{GRB~060418}

The {\it Swift} BAT instrument
triggered on GRB~060418 at UT 2006 April 18 03:06:08 \citep{gcn4966}.  
The satellite reported an immediate XRT position and 
\cite{gcn4966} reported a bright, optical afterglow at 
RA=$15^h 45^m 42^s.8$, DEC=$-03^\circ 38' 26''.1$.  We began observations shortly after
this announcement.

\begin{figure}
\epsscale{0.85}
\plotone{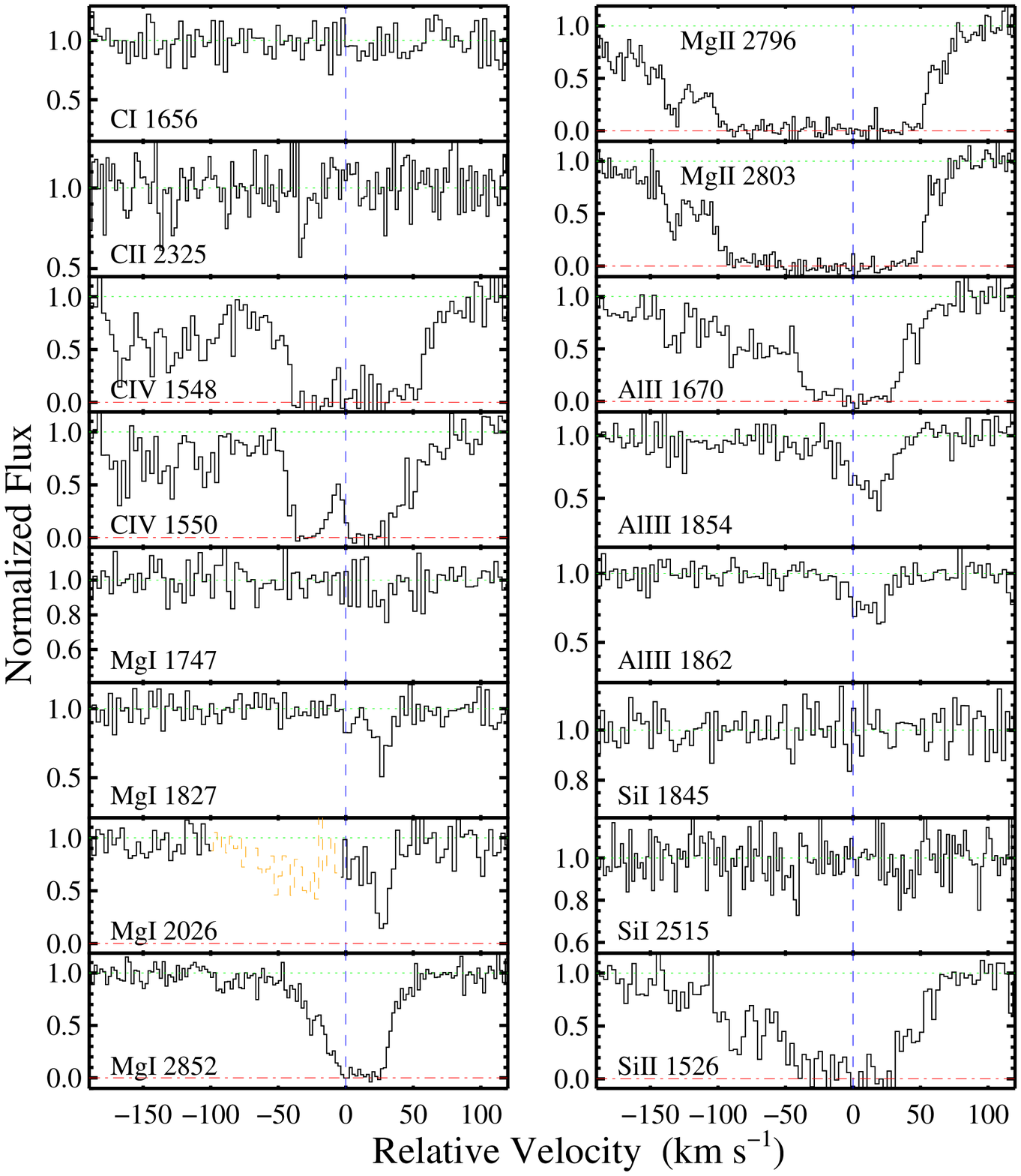}
\caption{Velocity profiles of the resonance transitions identified in the 
ISM of the host galaxy of GRB~060418.  The velocity $v=0\mkms$
corresponds to $z = 1.490$.  The profiles are ordered by atomic
number and ionization state.
}
\label{fig:060418res}
\end{figure}

\begin{figure}
\epsscale{0.85}
\plotone{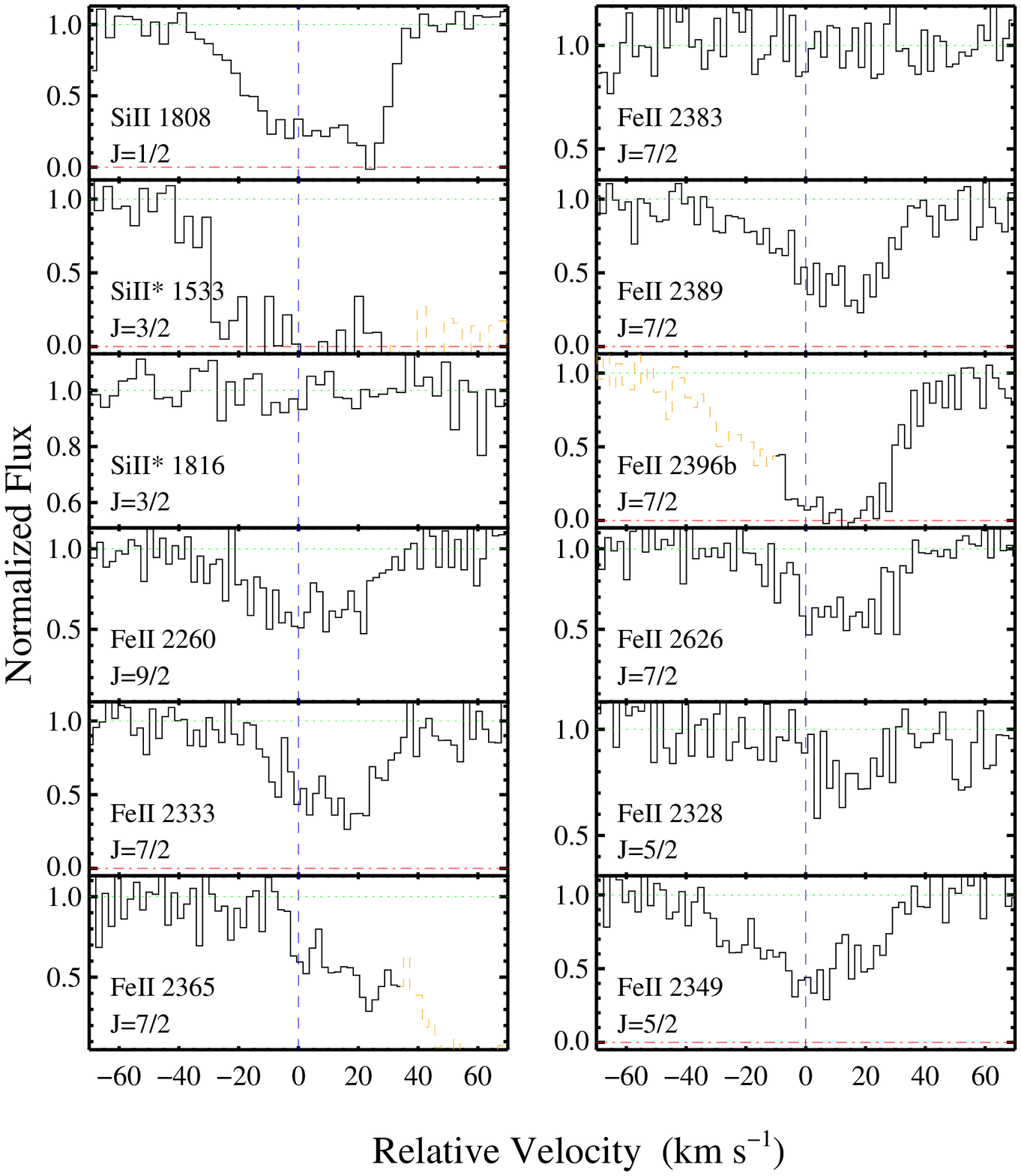}
\caption{Velocity profiles of the fine-structure transitions 
(with several resonance lines for comparison) identified
in the ISM of the host galaxy of GRB~060418.  The velocity $v=0\mkms$
corresponds to $z = 1.490$.  The profiles are ordered by atomic
number and energy level.
}
\label{fig:060418fine}
\end{figure}

Figure~\ref{fig:060418res} shows the velocity profiles for
the resonance lines associated with the ISM of the host galaxy.
Figure~\ref{fig:060418fine} shows the fine-structure transitions
and a resonance line for comparison.  Finally, Table~\ref{tab:060418}
lists the column density measurements for all of the
transitions.

\subsection{Summary}
\label{sec:summ}

Table~\ref{tab:summ} presents a summary of the gas-phase abundances
observed along the four sightlines.  In those cases where
the $\mnhi$ value was not measured (GRB~051111, GRB~060418), 
we have assumed a conservative
upper limit of $10^{23} \cm{-2}$.  
The relative gas-phase abundances [X/Fe] and [X/$\alpha$] assume solar
relative abundances \citep{solabnd} and do not include ionization
corrections nor differential depletion corrections.

\section{Strong, Intervening \ion{Mg}{2} Absorbers}
\label{sec:mgii}

\begin{figure}
\epsscale{0.85}
\plotone{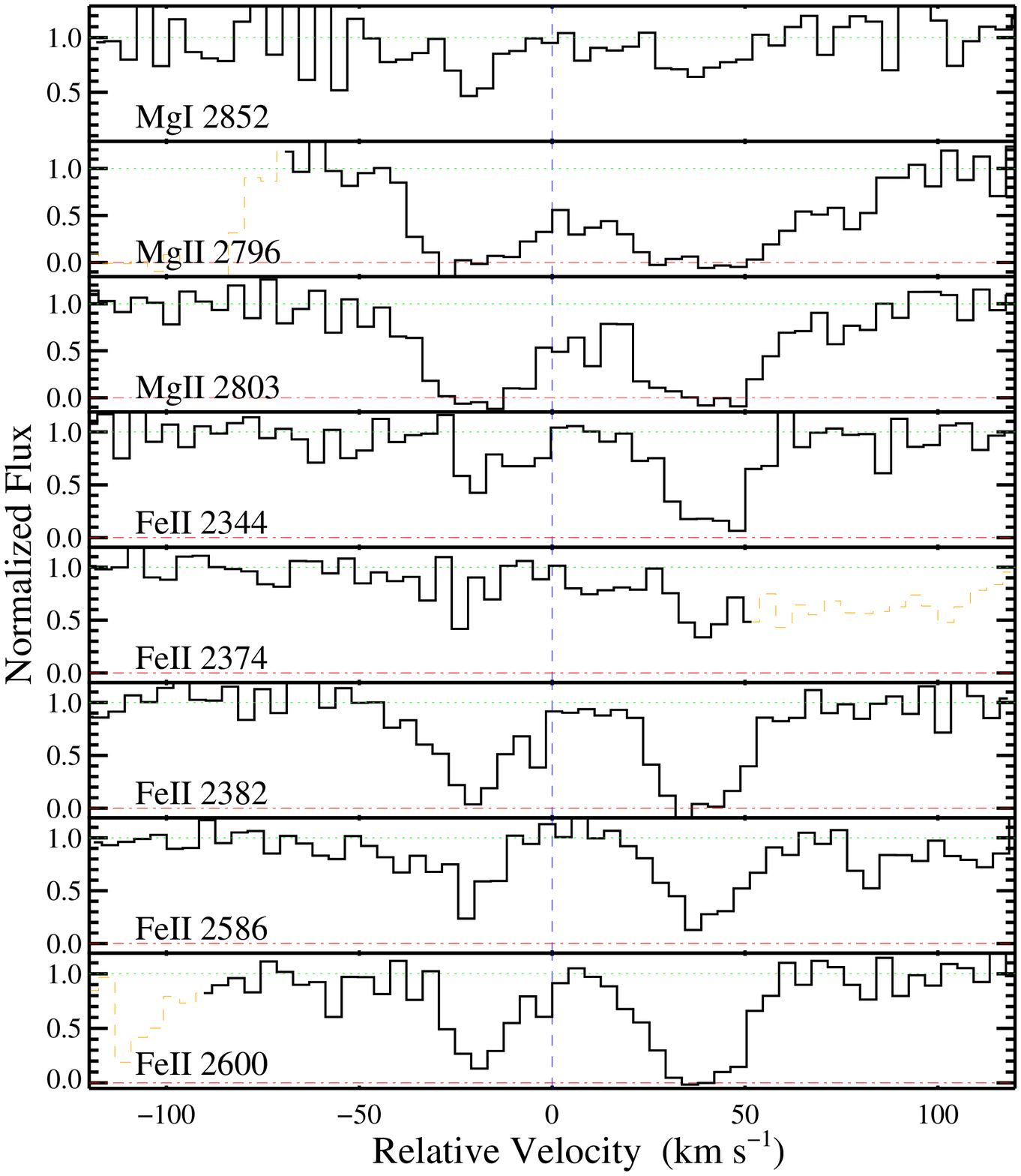}
\caption{Velocity profiles of the transitions identified
with the \ion{Mg}{2} absorber at $z=1.773$ toward
GRB~050730.  
}
\label{fig:050730Mg1}
\end{figure}

\begin{figure}
\epsscale{0.85}
\plotone{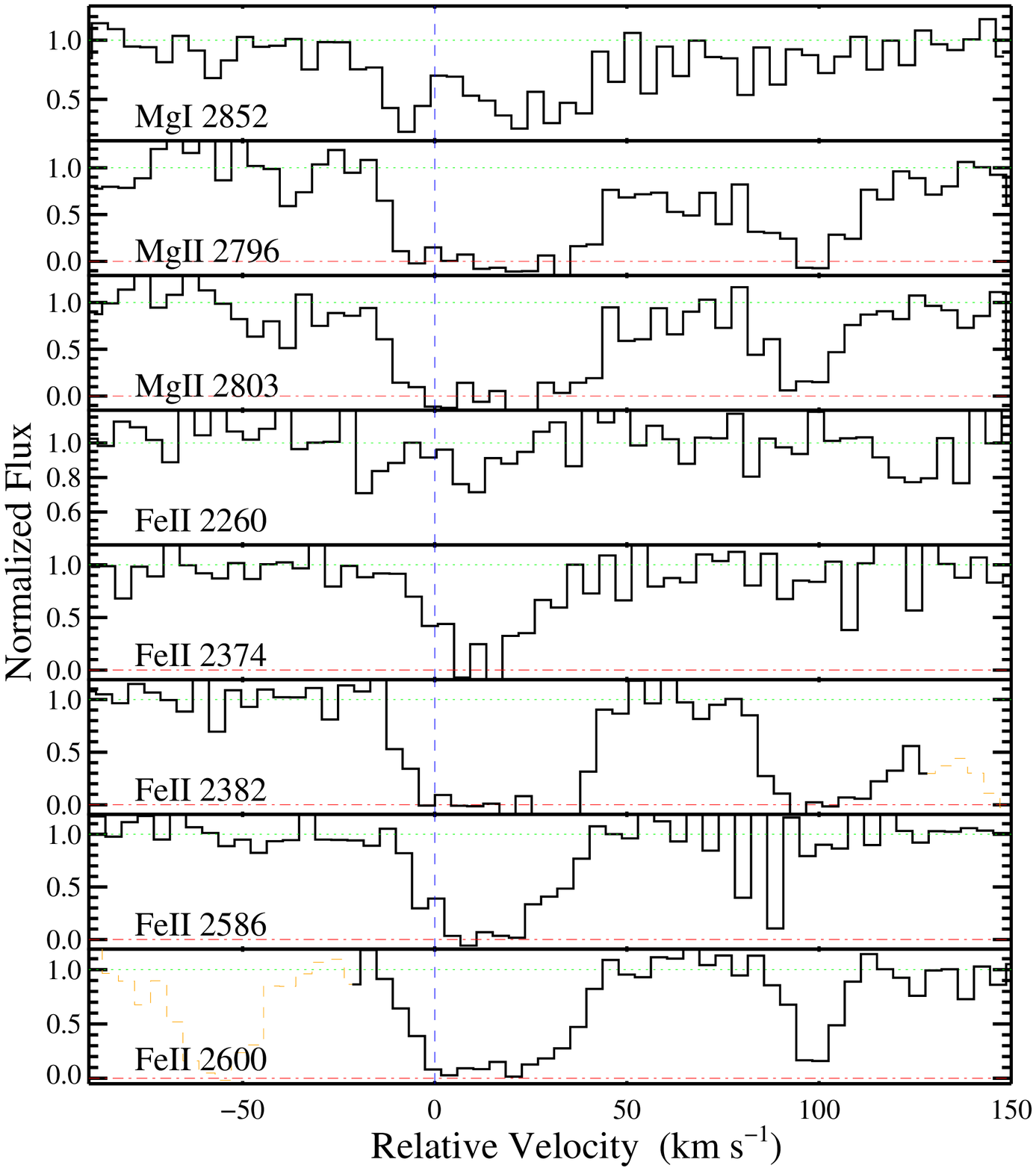}
\caption{Velocity profiles of the transitions identified
with the \ion{Mg}{2} absorber at $z=2.253$ toward
GRB~050730.  
}
\label{fig:050730Mg2}
\end{figure}

\begin{figure}
\epsscale{0.85}
\plotone{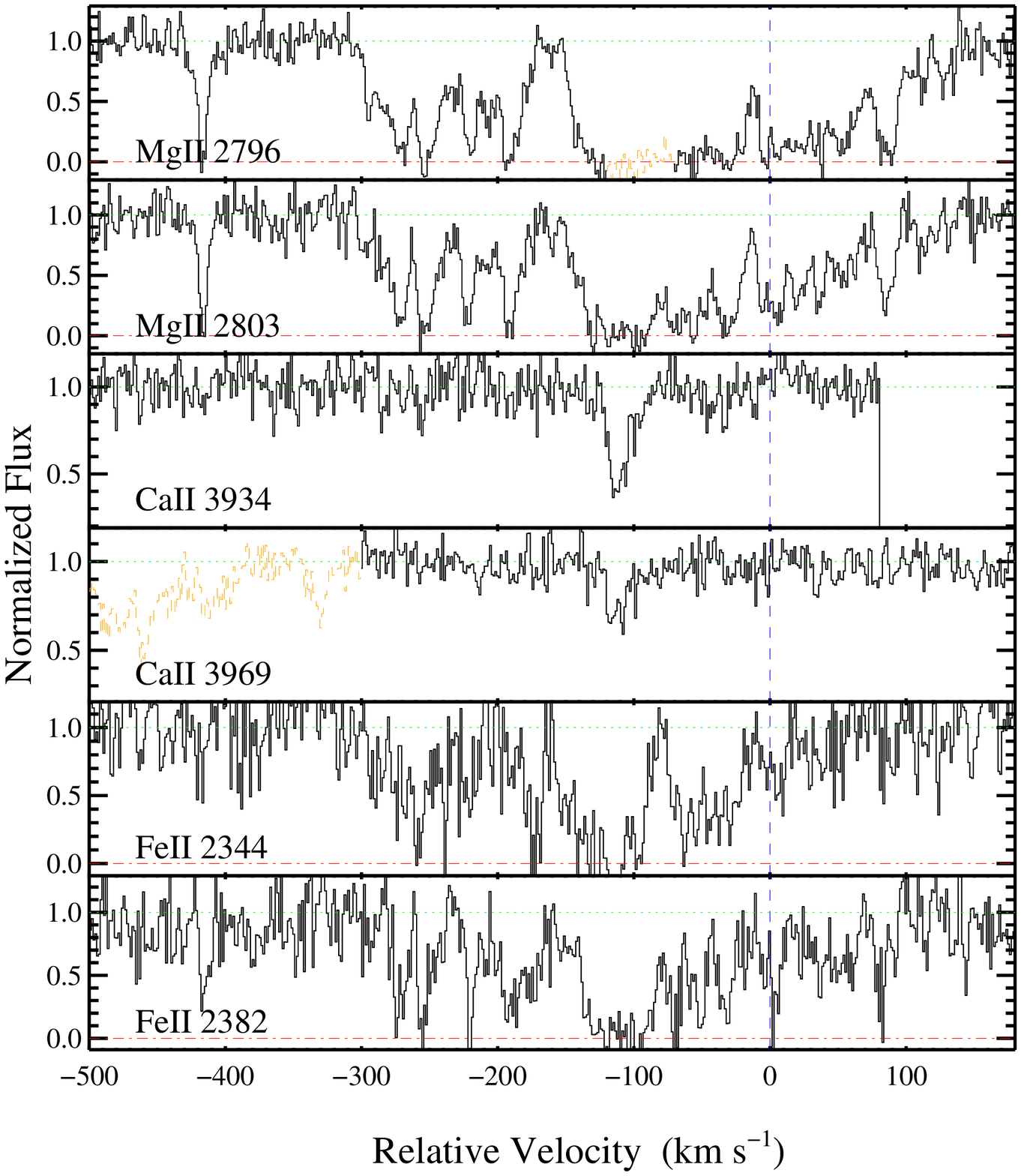}
\caption{Velocity profiles of the transitions identified
with the \ion{Mg}{2} absorber at $z=0.692$ toward
GRB~050820.  
}
\label{fig:050820Mg1}
\end{figure}

\begin{figure}
\epsscale{0.85}
\plotone{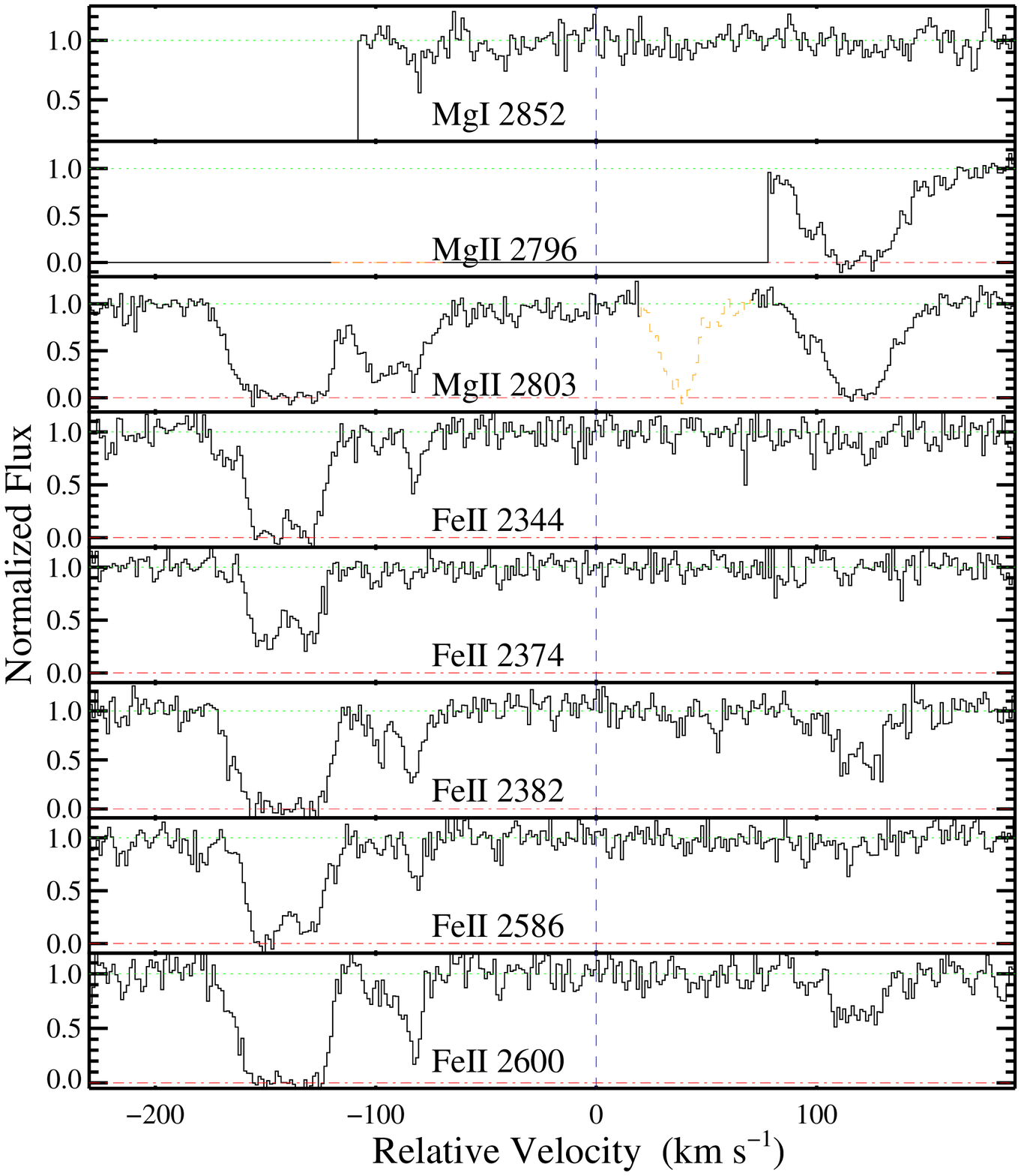}
\caption{Velocity profiles of the transitions identified
with the \ion{Mg}{2} absorber at $z=1.430$ toward
GRB~050820.  
}
\label{fig:050820Mg2}
\end{figure}

\begin{figure}
\epsscale{0.85}
\plotone{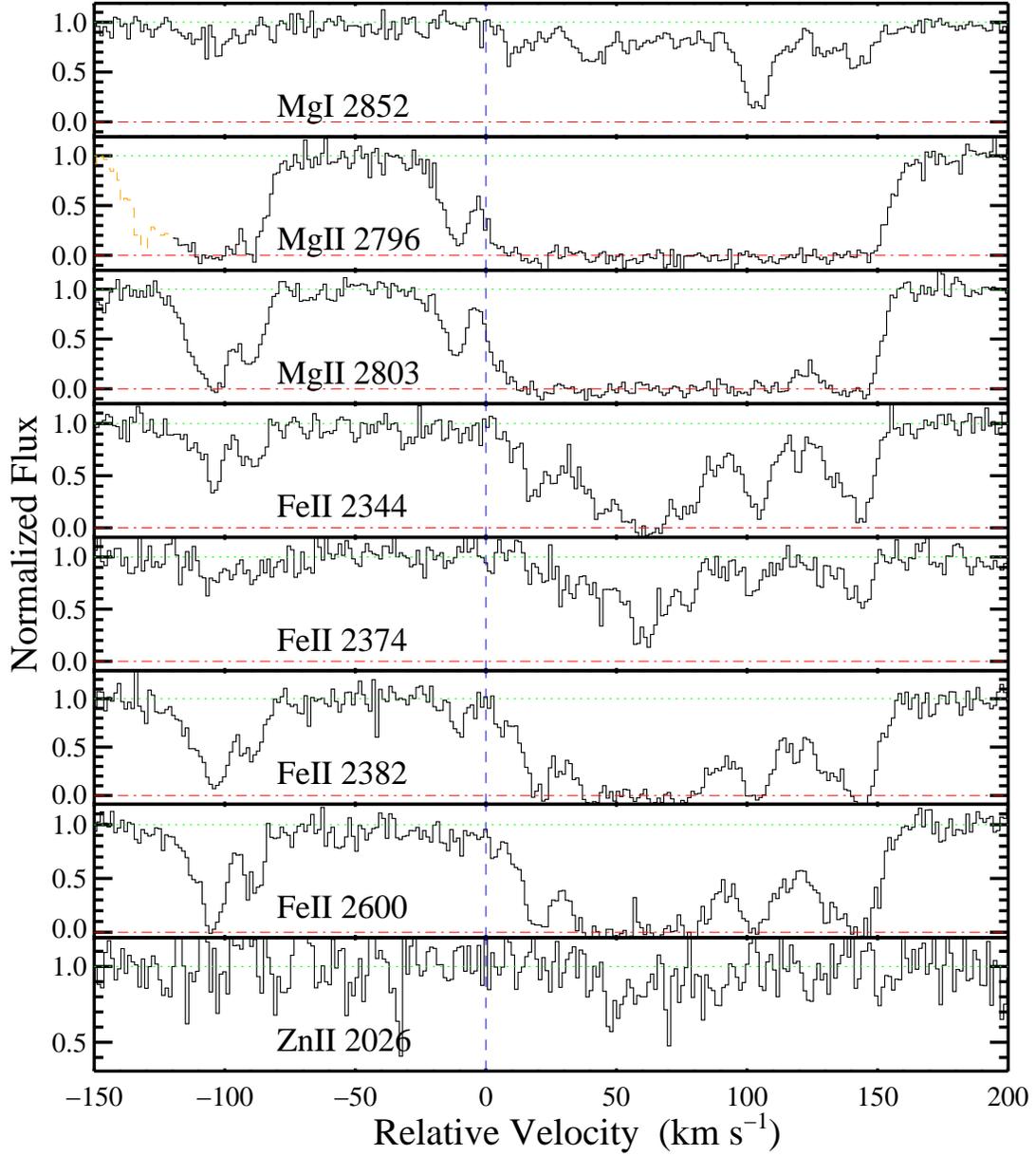}
\caption{Velocity profiles of the transitions identified
with the \ion{Mg}{2} absorber at $z=1.189$ toward
GRB~051111.  
}
\label{fig:051111Mg1}
\end{figure}

\begin{figure}
\epsscale{0.85}
\plotone{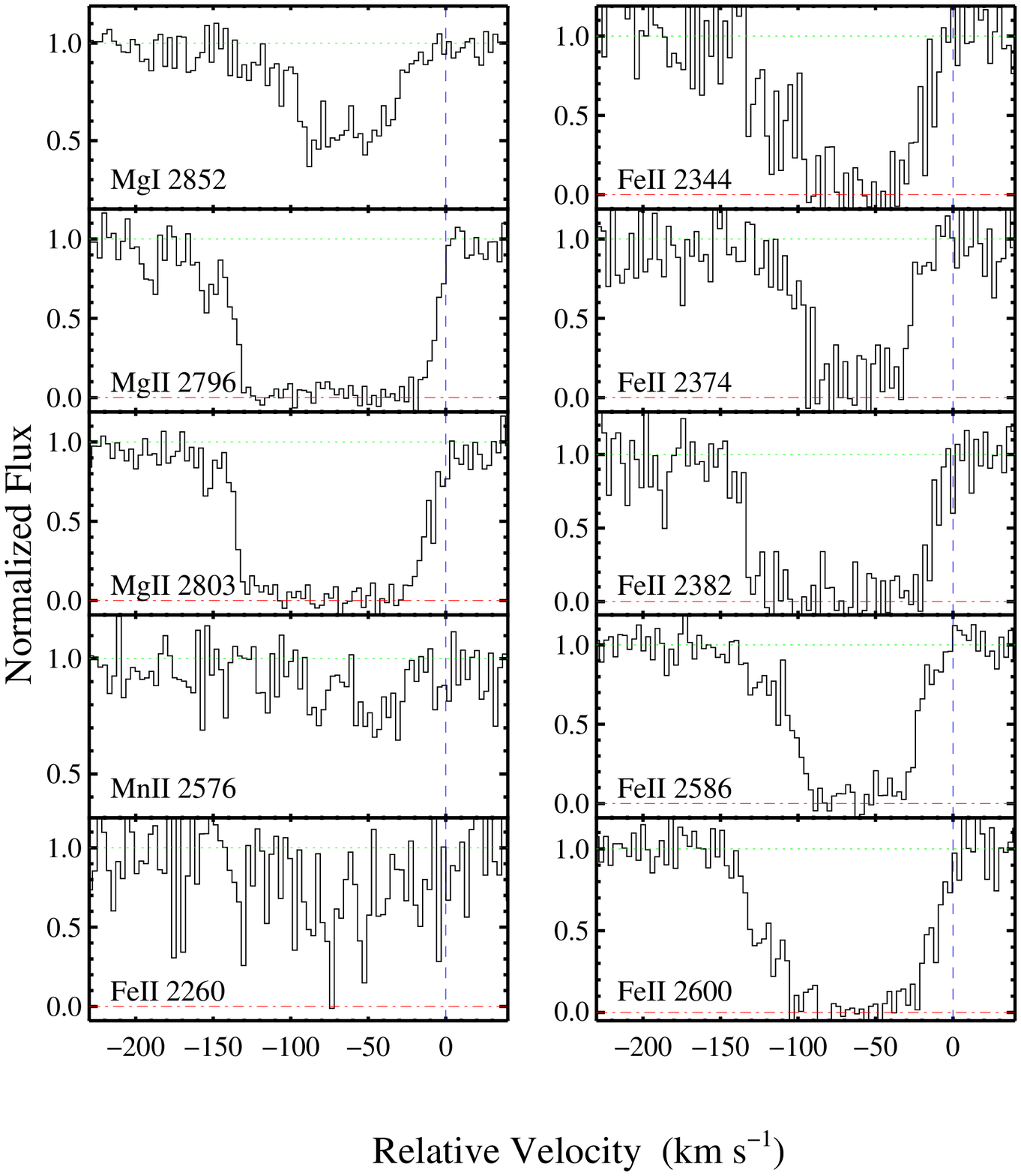}
\caption{Velocity profiles of the transitions identified
with the \ion{Mg}{2} absorber at $z=0.603$ toward
GRB~060418.  
}
\label{fig:060418Mg1}
\end{figure}

\begin{figure}
\epsscale{0.85}
\plotone{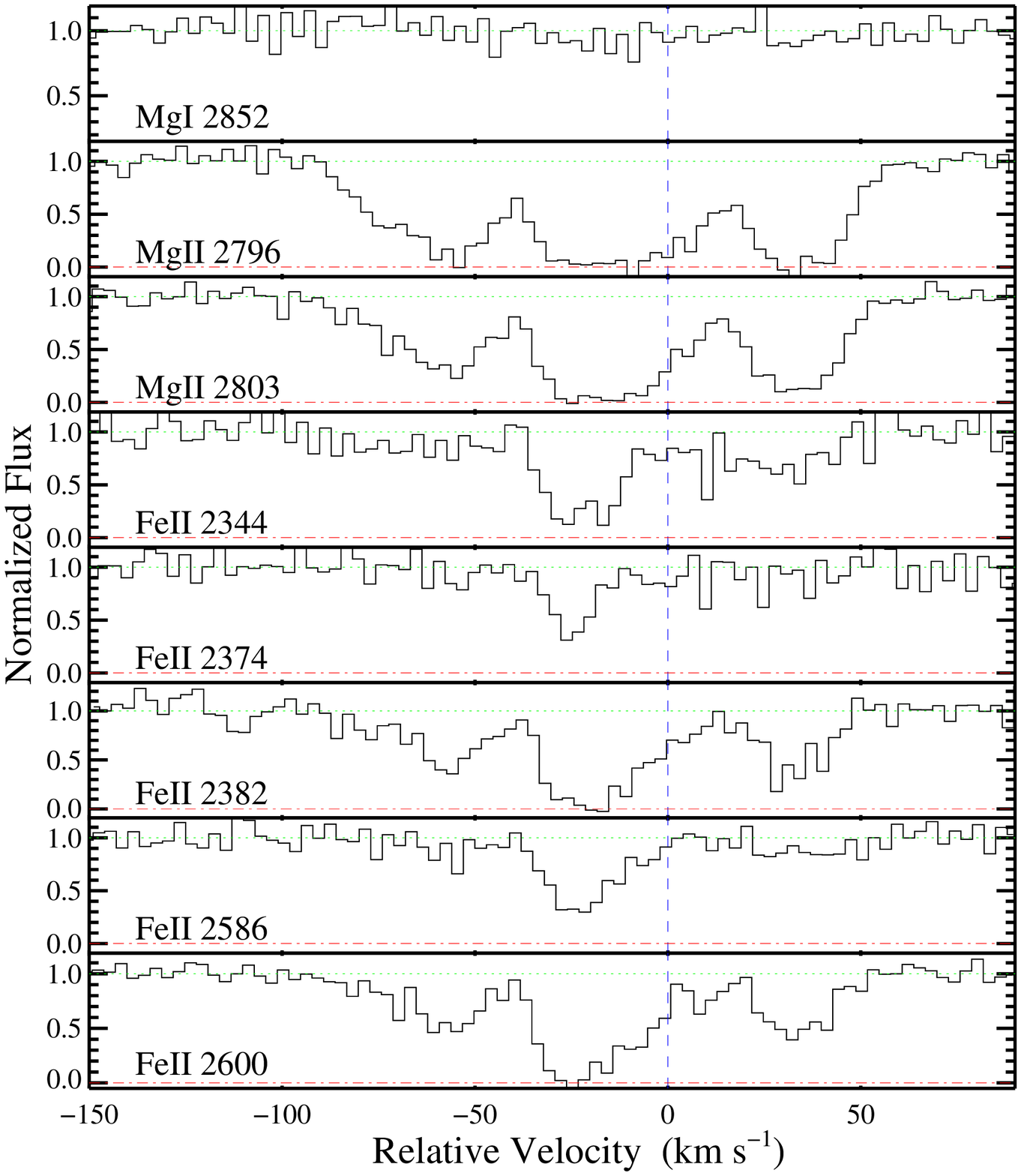}
\caption{Velocity profiles of the transitions identified
with the \ion{Mg}{2} absorber at $z=0.656$ toward
GRB~060418.  
}
\label{fig:060418Mg2}
\end{figure}

\begin{figure}
\epsscale{0.85}
\plotone{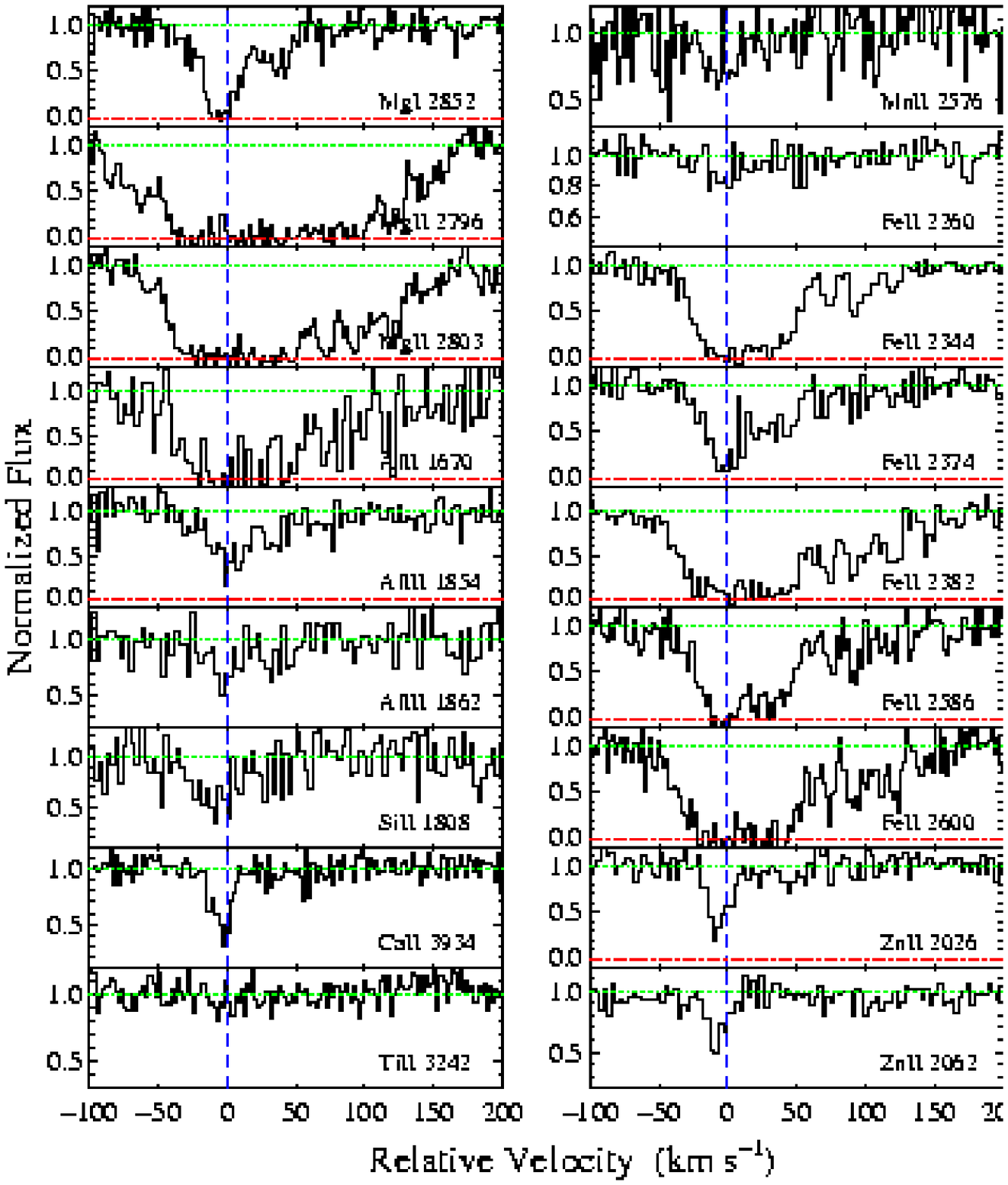}
\caption{Velocity profiles of the transitions identified
with the \ion{Mg}{2} absorber at $z=1.107$ toward
GRB~060418.  
}
\label{fig:060418Mg3}
\end{figure}

In a companion paper \citep{ppc+06}, we have shown that there
is an enhancement of strong \ion{Mg}{2} absorbers along GRB sightlines
in comparison with QSO sightlines.  Here, we present the full set of
velocity profiles and column density measurements for the absorbers along
the four GRB sightlines considered in that paper.
In Figures~\ref{fig:050730Mg1}-\ref{fig:060418Mg3} we present
the \ion{Mg}{2} absorption systems with $W_r > 0.5$\AA. 
The ionic column densities
and equivalent widths are reported in Tables~\ref{tab:050730MgII1}-\ref{tab:060418MgII3}.
We have measured the $D$-statistics \citep{ellison06} for each profile and
find $D>6.8$ in every case.  This suggests that a significant fraction
$(>50\%$) of the absorbers are damped \lya\ systems.
In Table~\ref{tab:mgsumm} we present a summary of these quantities.
This includes the minimum \nhi\ value implied by the observed ionic
column densities.  In general, the values further support the conclusion
that the majority of these absorbers are DLA systems.

\acknowledgments

The authors wish to recognize and acknowledge the very significant
cultural role and reverence that the summit of Mauna Kea has always
had within the indigenous Hawaiian community.  We are most fortunate
to have the opportunity to conduct observations from this mountain.
We acknowledge the efforts of S. Vogt, G. Marcy, J. Wright and
K. Hurley in obtaining the observations of GRB~050820.
J.X.P. is partially supported by NASA/Swift grant NNG05GF55G.


\clearpage

\clearpage
\begin{deluxetable}{lcccccccccc}
\rotate
\tablewidth{0pc}
\tablecaption{OBSERVATION LOG\label{tab:obslog}}
\tabletypesize{\scriptsize}
\tablehead{\colhead{GRB} & \colhead{UT$_{\rm burst}$} & \colhead{$z_{GRB}$} & 
\colhead{Telescope} & \colhead{Instr} & \colhead{FWHM} & \colhead{UT$_{start}$} 
& \colhead{UT$_{end}$} & \colhead{AM$^a$} & \colhead{$\lambda$} & \colhead{SNR$^b$} \\ 
& & & & & (\kms) & & & & (\AA)} 
\startdata
050730 & 19:58:23 & 3.969 & Magellan/Clay & MIKE   & 12 & 2005 Jul 31 00:00:04 & 2005 Jul 31 00:30:04 & 1.23   & 3400-9400 & 11 \\
       &          &       &               &        & 12 & 2005 Jul 31 00:32:52 & 2005 Jul 31 01:02:52 & 1.33   & 3400-9400 & 10 \\
       &          &       &               &        & 12 & 2005 Jul 31 01:05:11 & 2005 Jul 31 01:35:11 & 1.48   & 3400-9400 &  8 \\
050820 & 06:34:53 & 2.615 & Keck\,I       & HIRESr &  7.5 & 2005 Aug 20 07:28:59 & 2005 Aug 20 07:43:59 & 1.51 & 3800-8000 & 14 \\
       &          &       &               &        &  7.5 & 2005 Aug 20 07:44:42 & 2005 Aug 20 07:59:43 & 1.41 & 3800-8000 & 12 \\
051111 & 05:59:39 & 1.549 & Keck\,I       & HIRESr &  5 & 2005 Aug 11 07:02:58 & 2005 Aug 11 07:32:59 & 1.02   & 4160-8720 & 22 \\
       &          &       &               &        &  5 & 2005 Aug 11 07:34:23 & 2005 Aug 11 08:04:23 & 1.06   & 4160-8720 & 18 \\
       &          &       &               &        &  5 & 2005 Aug 11 08:05:22 & 2005 Aug 11 08:35:23 & 1.12   & 4160-8720 & 14 \\
060418 & 03:06:08 & 4.048 & Magellan/Clay & MIKE   & 12 & 2006 Apr 18 03:34:02 & 2006 Apr 18 03:54:02 & 1.61   & 3400-9400 & 21 \\
       &          &       &               &        & 12 & 2006 Apr 18 04:01:08 & 2006 Apr 18 04:31:08 & 1.44   & 3400-9400 & 17 \\
       &          &       &               &        & 12 & 2006 Apr 18 04:54:31 & 2006 Apr 18 05:24:31 & 1.24   & 3400-9400 & 9 \\
\enddata
\tablenotetext{a}{Airmass at the start of the observation.}
\tablenotetext{b}{Signal-to-noise ratio per resolution element (FWHM) near the center of the
spectral range.}
\end{deluxetable}

\begin{deluxetable}{lccccccccc}
\tablewidth{0pc}
\tablecaption{IONIC COLUMN DENSITIES FOR HG~050730\label{tab:050730}}
\tabletypesize{\scriptsize}
\tablehead{\colhead{Ion} & \colhead{$J^a$} & \colhead{$E_{J}$} & \colhead{$\lambda$} & \colhead{$\log f$}
& \colhead{$v_{int}^b$} 
& \colhead{$W_\lambda^c$} & \colhead{$\log N$} & \colhead{$\log N_{adopt}$} \\
& & (cm$^{-1}$) & (\AA) & & (\kms) & (m\AA) }
\startdata
\ion{C}{1}\\
&0 &    0.00 & 1277.2450 &$ -1.0148$&$[ -40,  40]$&$<  20.0$&$<13.37$&$< 13.37$\\
&0 &    0.00 & 1656.9283 &$ -0.8523$&$[ -30,  50]$&$<  20.4$&$<13.47$&\\
\ion{C}{2}\\
&1/2 &    0.00 & 1334.5323 &$ -0.8935$&$[-150,  50]$&$ 577.3 \pm  11.4$&$>14.95$&$> 14.95$\\
&3/2 &   63.40 & 1335.7077 &$ -0.9397$&$[-100,  50]$&$ 452.4 \pm   9.9$&$>14.85$&$> 14.85$\\
\ion{C}{4}\\
& &    0.00 & 1548.1950 &$ -0.7194$&$[-150,  70]$&$ 807.3 \pm  13.7$&$>14.79$&$> 15.02$\\
& &    0.00 & 1550.7700 &$ -1.0213$&$[-150,  70]$&$ 685.1 \pm  16.0$&$>15.02$&\\
\ion{N}{1}\\
& &    0.00 & 1134.1653 &$ -1.8722$&$[ -30,  30]$&$  67.6 \pm   7.3$&$14.86 \pm 0.08$&$ 14.78 \pm 0.05$\\
& &    0.00 & 1134.4149 &$ -1.5714$&$[ -30,  30]$&$ 106.4 \pm   8.4$&$14.75 \pm 0.07$&\\
\ion{N}{5}\\
& &    0.00 & 1238.8210 &$ -0.8041$&$[ -40,  80]$&$ 142.1 \pm  21.3$&$14.03 \pm 0.10$&$ 14.09 \pm 0.08$\\
& &    0.00 & 1242.8040 &$ -1.1066$&$[ -40,  80]$&$  94.1 \pm  21.3$&$14.26 \pm 0.11$&\\
\ion{O}{1}\\
&2 &    0.00 & 1302.1685 &$ -1.3110$&$[ -90,  60]$&$ 412.9 \pm  10.3$&$>15.27$&$> 15.27$\\
&2 &    0.00 & 1355.5977 &$ -5.9066$&$[ -40,  40]$&$<  16.6$&$<18.09$&\\
&1 &  158.26 & 1304.8576 &$ -1.3118$&$[ -60,  50]$&$ 228.8 \pm   8.7$&$>14.94$&$> 14.94$\\
&0 &  226.98 & 1306.0286 &$ -1.3122$&$[ -60,  50]$&$ 125.0 \pm   9.2$&$>14.61$&$> 14.61$\\
\ion{Mg}{2}\\
& &    0.00 & 1239.9253 &$ -3.1986$&$[ -40,  40]$&$<  35.4$&$<16.08$&$< 16.08$\\
& &    0.00 & 1240.3947 &$ -3.4498$&$[ -40,  40]$&$<  34.7$&$<16.15$&\\
\ion{Al}{2}\\
& &    0.00 & 1670.7874 &$  0.2742$&$[-110,  50]$&$ 590.1 \pm  20.7$&$>13.54$&$> 13.54$\\
\ion{Si}{2}\\
&1/2 &    0.00 & 1260.4221 &$  0.0030$&$[-150,  60]$&$ 592.1 \pm  14.5$&$>14.08$&$> 14.69$\\
&1/2 &    0.00 & 1304.3702 &$ -1.0269$&$[ -90,  50]$&$ 291.5 \pm   9.8$&$>14.69$&\\
&1/2 &    0.00 & 1526.7066 &$ -0.8962$&$[-130,  50]$&$ 365.5 \pm  13.9$&$>14.55$&\\
&1/2 &    0.00 & 1808.0126 &$ -2.6603$&$[ -60,  40]$&$<  30.6$&$<15.59$&\\
&3/2 &  287.24 & 1264.7377 &$ -0.0441$&$[-100,  40]$&$ 469.6 \pm  13.3$&$>14.03$&$> 14.56$\\
&3/2 &  287.24 & 1265.0020 &$ -0.9983$&$[ -30,  40]$&$ 184.4 \pm  11.1$&$>14.56$&\\
&3/2 &  287.24 & 1309.2757 &$ -0.8333$&$[ -75,  50]$&$ 205.2 \pm  11.1$&$>14.28$&\\
&3/2 &  287.24 & 1817.4512 &$ -3.8894$&$[ -40,  40]$&$<  24.8$&$<16.03$&\\
\ion{Si}{4}\\
& &    0.00 & 1393.7550 &$ -0.2774$&$[-150,  50]$&$ 545.8 \pm  16.1$&$>14.21$&$> 14.52$\\
& &    0.00 & 1402.7700 &$ -0.5817$&$[-150,  50]$&$ 505.9 \pm  13.6$&$>14.52$&\\
\ion{P}{2}\\
&1/2 &    0.00 & 1301.8740 &$ -1.7620$&$[ -40,  20]$&$<  14.6$&$<14.23$&$< 14.23$\\
\ion{S}{2}\\
& &    0.00 & 1250.5840 &$ -2.2634$&$[ -30,  55]$&$  58.6 \pm  14.7$&$15.09 \pm 0.10$&$ 15.22 \pm 0.06$\\
& &    0.00 & 1253.8110 &$ -1.9634$&$[ -30,  55]$&$ 175.4 \pm  10.9$&$15.39 \pm 0.07$&\\
& &    0.00 & 1259.5190 &$ -1.7894$&$[ -30,  55]$&$ 129.2 \pm  11.1$&$>15.07$&\\
\ion{Ar}{1}\\
& &    0.00 & 1066.6600 &$ -1.1709$&$[ -30,  30]$&$  90.2 \pm   8.2$&$>14.38$&$> 14.38$\\
\ion{Fe}{2}\\
&9/2 &    0.00 & 1608.4511 &$ -1.2366$&$[-100,  50]$&$ 287.0 \pm  10.7$&$>14.72$&$ 14.98 \pm 0.12$\\
&9/2 &    0.00 & 1611.2005 &$ -2.8665$&$[ -30,  30]$&$  24.2 \pm   7.6$&$14.98 \pm 0.12$&\\
&7/2 &  384.79 & 1618.4680 &$ -1.6696$&$[ -30,  30]$&$  65.4 \pm   8.3$&$14.26 \pm 0.06$&$ 14.29 \pm 0.04$\\
&7/2 &  384.79 & 1621.6856 &$ -1.4191$&$[ -30,  30]$&$ 111.2 \pm   9.0$&$14.34 \pm 0.07$&\\
&5/2 &  667.68 & 1629.1596 &$ -1.4353$&$[ -30,  30]$&$  78.6 \pm  11.6$&$14.12 \pm 0.07$&$ 14.12 \pm 0.07$\\
&3/2 &  862.61 & 1634.3498 &$ -1.6882$&$[ -30,  30]$&$  30.3 \pm   9.4$&$13.95 \pm 0.11$&$ 13.92 \pm 0.06$\\
&3/2 &  862.61 & 1636.3313 &$ -1.3925$&$[ -30,  30]$&$  57.7 \pm   9.0$&$13.91 \pm 0.07$&\\
&1/2 &  977.05 & 1639.4012 &$ -1.2373$&$[ -30,  30]$&$  48.5 \pm   8.3$&$13.65 \pm 0.07$&$ 13.65 \pm 0.07$\\
\ion{Ni}{2}\\
& &    0.00 & 1317.2170 &$ -1.2366$&$[ -40,  40]$&$<  21.0$&$13.68 \pm 0.13$&$ 13.68 \pm 0.06$\\
& &    0.00 & 1370.1310 &$ -1.1141$&$[ -40,  50]$&$  46.8 \pm  10.7$&$13.68 \pm 0.09$&\\
& &    0.00 & 1741.5490 &$ -1.3696$&$[ -40,  40]$&$  43.2 \pm  12.3$&$13.69 \pm 0.11$&\\
\enddata
\tablenotetext{a}{Total angular momentum of the electron spin and orbital angular moment.
$E_{J}$ is the energy above the ground state.}
\tablenotetext{b}{Velocity interval over which the equivalent width and
column density are measured.}
\tablenotetext{c}{Rest equivalent width.}
\end{deluxetable}

\begin{deluxetable}{lccccccccc}
\tablewidth{0pc}
\tablecaption{IONIC COLUMN DENSITIES FOR HG~050820\label{tab:050820}}
\tabletypesize{\scriptsize}
\tablehead{\colhead{Ion} & \colhead{$J^a$} & \colhead{$E_{J}$} & \colhead{$\lambda$} & \colhead{$\log f$}
& \colhead{$v_{int}^b$} 
& \colhead{$W_\lambda^c$} & \colhead{$\log N$} & \colhead{$\log N_{adopt}$} \\
& & (cm$^{-1}$) & (\AA) & & (\kms) & (m\AA) }
\startdata
\ion{C}{1}\\
&0 &    0.00 & 1277.2450 &$ -1.0148$&$[ -40,  40]$&$<  12.1$&$<13.13$&$< 12.75$\\
&0 &    0.00 & 1560.3092 &$ -1.0947$&$[ -40,  40]$&$<  10.7$&$<12.96$&\\
&0 &    0.00 & 1656.9283 &$ -0.8523$&$[ -30,  50]$&$<  12.3$&$<12.75$&\\
\ion{C}{2}\\
&1/2 &    0.00 & 1334.5323 &$ -0.8935$&$[-220, 280]$&$2047.3 \pm   9.4$&$>15.54$&$> 15.54$\\
&3/2 &   63.40 & 1335.7077 &$ -0.9397$&$[  40, 100]$&$  30.3 \pm   4.4$&$13.28 \pm 0.06$&$ 13.28 \pm 0.06$\\
\ion{C}{4}\\
& &    0.00 & 1548.1950 &$ -0.7194$&$[-300, 200]$&$1505.3 \pm  10.4$&$>14.88$&$> 15.00$\\
& &    0.00 & 1550.7700 &$ -1.0213$&$[-300, 200]$&$1138.6 \pm  12.1$&$>15.00$&\\
\ion{N}{1}\\
& &    0.00 & 1159.8170 &$ -5.0701$&$[ -50,  50]$&$<  16.4$&$<18.26$&$< 18.26$\\
& &    0.00 & 1160.9370 &$ -5.6198$&$[ -50,  50]$&$<  14.6$&$<18.60$&\\
\ion{N}{2}\\
&1/2 &    0.00 & 1083.9900 &$ -0.9867$&$[-250, 200]$&$1084.8 \pm  24.7$&$>15.36$&$> 15.36$\\
\ion{N}{5}\\
& &    0.00 & 1238.8210 &$ -0.8041$&$[ -50,  50]$&$<  13.9$&$<13.00$&$< 13.00$\\
& &    0.00 & 1242.8040 &$ -1.1066$&$[ -40,  80]$&$<  15.1$&$<13.35$&\\
\ion{O}{1}\\
&2 &    0.00 & 1302.1685 &$ -1.3110$&$[-210, 280]$&$1562.1 \pm  10.9$&$>15.84$&$> 15.84$\\
&0 &  226.98 & 1306.0286 &$ -1.3122$&$[ -50,  50]$&$<  10.9$&$<13.41$&$< 13.41$\\
\ion{Mg}{1}\\
& &    0.00 & 1827.9351 &$ -1.6216$&$[ -40,  40]$&$<  10.9$&$<13.37$&$< 12.60$\\
& &    0.00 & 2026.4768 &$ -0.9508$&$[ -20,  40]$&$<  10.7$&$<12.60$&\\
\ion{Mg}{2}\\
& &    0.00 & 1239.9253 &$ -3.1986$&$[ -40,  40]$&$  48.1 \pm   5.9$&$15.86 \pm 0.05$&$ 15.86 \pm 0.05$\\
& &    0.00 & 1240.3947 &$ -3.4498$&$[ -40,  40]$&$<  10.2$&$<17.13$&\\
\ion{Al}{2}\\
& &    0.00 & 1670.7874 &$  0.2742$&$[-220, 280]$&$1810.8 \pm   9.9$&$>14.09$&$> 14.09$\\
\ion{Al}{3}\\
& &    0.00 & 1854.7164 &$ -0.2684$&$[-220, 280]$&$ 698.9 \pm  11.9$&$>13.88$&$ 13.90 \pm 0.01$\\
& &    0.00 & 1862.7895 &$ -0.5719$&$[-220, 280]$&$ 458.5 \pm  11.8$&$13.90 \pm 0.01$&\\
\ion{Si}{1}\\
& &    0.00 & 1845.5200 &$ -0.6402$&$[ -40,  40]$&$<  12.4$&$<12.44$&$< 12.44$\\
\ion{Si}{2}\\
&1/2 &    0.00 & 1260.4221 &$  0.0030$&$[-210, 280]$&$1703.3 \pm  12.2$&$>14.57$&$> 15.43$\\
&1/2 &    0.00 & 1304.3702 &$ -1.0269$&$[-210, 280]$&$1280.7 \pm  10.4$&$>15.43$&\\
&1/2 &    0.00 & 1526.7066 &$ -0.8962$&$[-220, 280]$&$1646.9 \pm  11.3$&$>15.28$&\\
&3/2 &  287.24 & 1264.7377 &$ -0.0441$&$[-210, 120]$&$ 673.5 \pm  10.6$&$>14.05$&$ 13.73 \pm 0.02$\\
&3/2 &  287.24 & 1265.0020 &$ -0.9983$&$[ -20,  40]$&$<   9.3$&$<13.83$&\\
&3/2 &  287.24 & 1533.4312 &$ -0.6396$&$[-190,  20]$&$ 198.6 \pm   8.3$&$13.73 \pm 0.02$&\\
&3/2 &  287.24 & 1817.4512 &$ -3.8894$&$[ -40,  40]$&$<  10.9$&$<15.64$&\\
\ion{Si}{3}\\
& &    0.00 & 1892.0300 &$ -4.5702$&$[ -40,  40]$&$<   9.5$&$<16.24$&$< 16.24$\\
\ion{Si}{4}\\
& &    0.00 & 1393.7550 &$ -0.2774$&$[-200, 200]$&$ 817.7 \pm  10.7$&$>14.23$&$> 14.34$\\
& &    0.00 & 1402.7700 &$ -0.5817$&$[-220, 200]$&$ 521.6 \pm  12.1$&$>14.34$&\\
\ion{P}{2}\\
&1/2 &    0.00 & 1152.8180 &$ -0.6271$&$[ -80,  30]$&$  86.6 \pm   8.0$&$13.64 \pm 0.04$&$ 13.64 \pm 0.04$\\
&1/2 &    0.00 & 1532.5330 &$ -2.1186$&$[ -60,  60]$&$<  13.4$&$<14.12$&\\
\ion{S}{2}\\
& &    0.00 & 1250.5840 &$ -2.2634$&$[-210, 140]$&$ 163.4 \pm  13.1$&$15.57 \pm 0.04$&$ 15.57 \pm 0.04$\\
& &    0.00 & 1253.8110 &$ -1.9634$&$[-210, 140]$&$ 364.6 \pm  12.0$&$>15.67$&\\
\ion{Ar}{1}\\
& &    0.00 & 1066.6600 &$ -1.1709$&$[ -50,  50]$&$ 135.1 \pm  15.7$&$>14.61$&$> 14.61$\\
\ion{Cr}{2}\\
& &    0.00 & 2056.2539 &$ -0.9788$&$[ -50,  50]$&$  69.8 \pm   5.5$&$13.31 \pm 0.03$&$ 13.33 \pm 0.02$\\
& &    0.00 & 2062.2340 &$ -1.1079$&$[ -50,  30]$&$  59.3 \pm   5.1$&$13.35 \pm 0.04$&\\
& &    0.00 & 2066.1610 &$ -1.2882$&$[ -40,  40]$&$  41.9 \pm   5.6$&$13.37 \pm 0.06$&\\
\ion{Fe}{2}\\
&9/2 &    0.00 & 1608.4511 &$ -1.2366$&$[-200, 150]$&$ 895.5 \pm   9.5$&$>15.15$&$ 14.82 \pm 0.12$\\
&9/2 &    0.00 & 1611.2005 &$ -2.8665$&$[ -80,  10]$&$  17.9 \pm   5.4$&$14.82 \pm 0.12$&\\
&7/2 &  384.79 & 1618.4680 &$ -1.6696$&$[ -30,  30]$&$<   9.2$&$<13.66$&$< 13.31$\\
&7/2 &  384.79 & 1621.6856 &$ -1.4191$&$[ -30,  30]$&$<   8.6$&$<13.31$&\\
&5/2 &  667.68 & 1629.1596 &$ -1.4353$&$[ -30,  30]$&$<  10.9$&$<13.29$&$< 13.29$\\
&3/2 &  862.61 & 1634.3498 &$ -1.6882$&$[ -30,  30]$&$<   9.2$&$<13.46$&$< 13.15$\\
&3/2 &  862.61 & 1636.3313 &$ -1.3925$&$[ -30,  30]$&$<   9.3$&$<13.15$&\\
&1/2 &  977.05 & 1639.4012 &$ -1.2373$&$[ -30,  30]$&$<   8.7$&$<12.97$&$< 12.97$\\
\ion{Ni}{2}\\
& &    0.00 & 1370.1310 &$ -1.1141$&$[ -90, 100]$&$<  16.2$&$<13.96$&$ 13.69 \pm 0.04$\\
& &    0.00 & 1703.4050 &$ -2.2218$&$[ -50,  50]$&$<  11.0$&$<14.04$&\\
& &    0.00 & 1709.6042 &$ -1.4895$&$[ -50,  50]$&$  37.0 \pm   5.9$&$13.70 \pm 0.07$&\\
& &    0.00 & 1741.5490 &$ -1.3696$&$[ -40,  40]$&$  43.6 \pm   5.5$&$13.66 \pm 0.05$&\\
& &    0.00 & 1751.9157 &$ -1.5575$&$[ -50,  50]$&$  35.5 \pm   5.8$&$13.73 \pm 0.07$&\\
\ion{Zn}{2}\\
& &    0.00 & 2026.1360 &$ -0.3107$&$[ -50,  50]$&$ 124.3 \pm   6.5$&$12.95 \pm 0.02$&$ 12.96 \pm 0.02$\\
& &    0.00 & 2062.6640 &$ -0.5918$&$[ -30,  50]$&$  78.3 \pm   5.0$&$12.98 \pm 0.03$&\\
\enddata
\tablenotetext{a}{Total angular momentum of the electron spin and orbital angular moment.
$E_{J}$ is the energy above the ground state.}
\tablenotetext{b}{Velocity interval over which the equivalent width and
column density are measured.}
\tablenotetext{c}{Rest equivalent width.}
\end{deluxetable}

\begin{deluxetable}{lccccccccc}
\tablewidth{0pc}
\tablecaption{IONIC COLUMN DENSITIES FOR HG~051111\label{tab:051111}}
\tabletypesize{\footnotesize}
\tablehead{\colhead{Ion} & \colhead{$J^a$} & \colhead{$E_{J}$} & \colhead{$\lambda$} & \colhead{$\log f$}
& \colhead{$v_{int}^b$} 
& \colhead{$W_\lambda^c$} & \colhead{$\log N$} & \colhead{$\log N_{adopt}$} \\
& & (cm$^{-1}$) & (\AA) & & (\kms) & (m\AA) }
\startdata
\ion{C}{1}\\
&0 &    0.00 & 1656.9283 &$ -0.8523$&$[ -40,  40]$&$<  20.6$&$<12.99$&$< 12.99$\\
\ion{C}{2}\\
&1/2 &    0.00 & 2325.4030 &$ -7.3484$&$[ -40,  40]$&$<  10.0$&$<18.85$&$< 18.84$\\
\ion{Mg}{1}\\
& &    0.00 & 1747.7937 &$ -2.0419$&$[ -40,  40]$&$  72.3 \pm   6.1$&$14.68 \pm 0.04$&$ 14.68 \pm 0.04$\\
& &    0.00 & 1827.9351 &$ -1.6216$&$[ -40,  40]$&$ 133.3 \pm   5.9$&$>14.63$&\\
& &    0.00 & 2026.4768 &$ -0.9508$&$[ -40,  40]$&$ 279.7 \pm   4.3$&$>14.31$&\\
\ion{Mg}{2}\\
& &    0.00 & 2796.3520 &$ -0.2130$&$[-200,  70]$&$2102.1 \pm   6.3$&$>14.35$&$> 14.62$\\
& &    0.00 & 2803.5310 &$ -0.5151$&$[-200,  70]$&$1994.4 \pm   6.7$&$>14.62$&\\
\ion{Al}{2}\\
& &    0.00 & 1670.7874 &$  0.2742$&$[-180,  80]$&$ 945.5 \pm  14.0$&$>13.75$&$> 13.75$\\
\ion{Al}{3}\\
& &    0.00 & 1854.7164 &$ -0.2684$&$[-130,  40]$&$ 331.1 \pm   8.7$&$>13.57$&$ 13.63 \pm 0.02$\\
& &    0.00 & 1862.7895 &$ -0.5719$&$[-130,  40]$&$ 245.3 \pm   7.8$&$13.63 \pm 0.02$&\\
\ion{Si}{1}\\
& &    0.00 & 1845.5200 &$ -0.6402$&$[ -40,  40]$&$<  12.2$&$<12.42$&$< 12.42$\\
& &    0.00 & 2515.0730 &$ -0.7905$&$[ -40,  40]$&$<  12.3$&$<12.43$&\\
\ion{Si}{2}\\
&1/2 &    0.00 & 1808.0130 &$ -2.6603$&$[-100,  40]$&$ 338.4 \pm   7.8$&$>16.14$&$> 16.14$\\
&1/2 &    0.00 & 2335.1230 &$ -5.3716$&$[ -40,  40]$&$<   9.3$&$<17.32$&\\
&3/2 &  287.24 & 1816.9285 &$ -2.7799$&$[ -40,  40]$&$  42.2 \pm   6.3$&$15.00 \pm 0.06$&$ 15.00 \pm 0.06$\\
\ion{Si}{3}\\
& &    0.00 & 1892.0300 &$ -4.5702$&$[ -40,  40]$&$<  11.1$&$<16.32$&$< 16.32$\\
\ion{Ti}{2}\\
&1/2 &    0.00 & 1910.6000 &$ -0.6946$&$[ -40,  40]$&$<  11.4$&$<12.42$&$< 12.42$\\
\ion{Cr}{2}\\
& &    0.00 & 2056.2539 &$ -0.9788$&$[ -40,  40]$&$ 154.9 \pm   5.1$&$13.90 \pm 0.03$&$ 13.88 \pm 0.01$\\
& &    0.00 & 2062.2340 &$ -1.1079$&$[ -40,  40]$&$ 149.4 \pm   5.4$&$13.89 \pm 0.02$&\\
& &    0.00 & 2066.1610 &$ -1.2882$&$[ -40,  40]$&$  88.8 \pm   5.5$&$13.85 \pm 0.03$&\\
\ion{Mn}{2}\\
& &    0.00 & 2576.8770 &$ -0.4549$&$[ -40,  40]$&$ 260.6 \pm   5.1$&$>13.55$&$> 13.64$\\
& &    0.00 & 2594.4990 &$ -0.5670$&$[ -40,  40]$&$ 229.5 \pm   4.1$&$>13.64$&\\
& &    0.00 & 2606.4620 &$ -0.7151$&$[ -40,  40]$&$ 213.2 \pm   5.3$&$>13.64$&\\
\ion{Fe}{1}\\
& &    0.00 & 2484.0210 &$ -0.2541$&$[ -40,  40]$&$<  11.3$&$<11.75$&$< 11.76$\\
\ion{Fe}{2}\\
&9/2 &    0.00 & 2249.8768 &$ -2.7397$&$[-100,  40]$&$ 112.9 \pm   6.8$&$15.28 \pm 0.02$&$ 15.32 \pm 0.01$\\
&9/2 &    0.00 & 2260.7805 &$ -2.6126$&$[-120,  40]$&$ 155.1 \pm   6.5$&$15.35 \pm 0.02$&\\
&9/2 &    0.00 & 2344.2140 &$ -0.9431$&$[-200,  80]$&$ 985.6 \pm   9.5$&$>14.70$&\\
&9/2 &    0.00 & 2374.4612 &$ -1.5045$&$[-180,  80]$&$ 569.0 \pm   7.9$&$>14.96$&\\
&9/2 &    0.00 & 2382.7650 &$ -0.4949$&$[-200,  80]$&$1313.9 \pm   8.6$&$>14.43$&\\
&9/2 &    0.00 & 2586.6499 &$ -1.1605$&$[-180, 100]$&$ 916.4 \pm   7.9$&$>14.76$&\\
&9/2 &    0.00 & 2600.1729 &$ -0.6216$&$[-200, 100]$&$1523.8 \pm   7.8$&$>14.59$&\\
&7/2 &  384.79 & 2333.5160 &$ -1.1601$&$[ -40,  40]$&$ 205.9 \pm   4.3$&$13.99 \pm 0.01$&$ 14.01 \pm 0.01$\\
&7/2 &  384.79 & 2365.5520 &$ -1.3054$&$[ -40,  40]$&$ 178.8 \pm   4.7$&$14.02 \pm 0.01$&\\
&7/2 &  384.79 & 2383.7884 &$ -2.2861$&$[ -40,  40]$&$  30.1 \pm   5.8$&$14.12 \pm 0.08$&\\
&7/2 &  384.79 & 2389.3582 &$ -1.0835$&$[ -40,  20]$&$ 216.9 \pm   5.3$&$>14.08$&\\
&7/2 &  384.79 & 2396.3559 &$ -0.5414$&$[ -40,  40]$&$ 405.8 \pm   4.9$&$>13.90$&\\
&7/2 &  384.79 & 2626.4511 &$ -1.3556$&$[ -40,  40]$&$ 182.3 \pm   5.2$&$14.01 \pm 0.01$&\\
&5/2 &  667.68 & 2328.1110 &$ -1.4492$&$[ -40,  40]$&$  85.6 \pm   4.6$&$13.78 \pm 0.02$&$ 13.78 \pm 0.01$\\
&5/2 &  667.68 & 2607.8664 &$ -0.9281$&$[ -40,  40]$&$ 232.3 \pm   5.5$&$13.78 \pm 0.02$&\\
&5/2 &  667.68 & 2618.3991 &$ -1.2967$&$[ -40,  40]$&$ 140.1 \pm   5.8$&$13.78 \pm 0.02$&\\
&3/2 &  862.61 & 2338.7250 &$ -1.0339$&$[ -40,  40]$&$ 126.8 \pm   4.8$&$13.57 \pm 0.02$&$ 13.59 \pm 0.01$\\
&3/2 &  862.61 & 2359.8280 &$ -1.2421$&$[ -40,  40]$&$  98.6 \pm   5.2$&$13.64 \pm 0.02$&\\
&3/2 &  862.61 & 2621.1912 &$ -2.4067$&$[ -40,  40]$&$<  11.5$&$<13.86$&\\
&1/2 &  977.05 & 2345.0010 &$ -0.8027$&$[ -25,  40]$&$  97.9 \pm   5.3$&$13.20 \pm 0.03$&$ 13.26 \pm 0.01$\\
&1/2 &  977.05 & 2414.0450 &$ -0.7557$&$[ -40,  40]$&$ 126.5 \pm   4.4$&$13.26 \pm 0.02$&\\
&1/2 &  977.05 & 2622.4518 &$ -1.2518$&$[ -40,  40]$&$  61.0 \pm   5.7$&$13.31 \pm 0.04$&\\
&1/2 &  977.05 & 2629.0777 &$ -0.7620$&$[ -40,  40]$&$ 156.3 \pm   4.9$&$13.29 \pm 0.02$&\\
\ion{Co}{2}\\
& &    0.00 & 1941.2852 &$ -1.4685$&$[ -40,  40]$&$<  10.3$&$<13.16$&$< 13.16$\\
\ion{Ni}{2}\\
& &    0.00 & 1703.4050 &$ -2.2218$&$[ -40,  40]$&$<  14.6$&$<14.46$&$ 13.97 \pm 0.03$\\
& &    0.00 & 1709.6042 &$ -1.4895$&$[ -40,  40]$&$  54.2 \pm   7.2$&$13.96 \pm 0.05$&\\
& &    0.00 & 1741.5531 &$ -1.3696$&$[ -40,  40]$&$  76.6 \pm   6.8$&$14.00 \pm 0.04$&\\
& &    0.00 & 1751.9157 &$ -1.5575$&$[ -40,  40]$&$  48.5 \pm   6.8$&$13.93 \pm 0.06$&\\
\ion{Zn}{2}\\
& &    0.00 & 2026.1360 &$ -0.3107$&$[ -40,  40]$&$ 237.9 \pm   4.4$&$>13.60$&$> 13.71$\\
& &    0.00 & 2062.6640 &$ -0.5918$&$[ -40,  40]$&$ 176.7 \pm   5.0$&$>13.71$&\\
\enddata
\tablenotetext{a}{Total angular momentum of the electron spin and orbital angular moment.
$E_{J}$ is the energy above the ground state.}
\tablenotetext{b}{Velocity interval over which the equivalent width and
column density are measured.}
\tablenotetext{c}{Rest equivalent width.}
\end{deluxetable}

\begin{deluxetable}{lccccccccc}
\tablewidth{0pc}
\tablecaption{IONIC COLUMN DENSITIES FOR HG~060418\label{tab:060418}}
\tabletypesize{\footnotesize}
\tablehead{\colhead{Ion} & \colhead{$J^a$} & \colhead{$E_{J}$} & \colhead{$\lambda$} & \colhead{$\log f$}
& \colhead{$v_{int}^b$} 
& \colhead{$W_\lambda^c$} & \colhead{$\log N$} & \colhead{$\log N_{adopt}$} \\
& & (cm$^{-1}$) & (\AA) & & (\kms) & (m\AA) }
\startdata
\ion{C}{1}\\
&0 &    0.00 & 1656.9283 &$ -0.8523$&$[ -50,  50]$&$<  18.2$&$<12.93$&$< 12.93$\\
\ion{C}{2}\\
&1/2 &    0.00 & 2325.4030 &$ -7.3484$&$[ -50,  50]$&$<  26.7$&$<19.28$&$< 19.28$\\
\ion{C}{4}\\
& &    0.00 & 1548.1950 &$ -0.7194$&$[-190, 110]$&$ 802.9 \pm  19.1$&$>14.68$&$> 14.88$\\
& &    0.00 & 1550.7700 &$ -1.0213$&$[-190, 110]$&$ 610.5 \pm  17.7$&$>14.88$&\\
\ion{Mg}{1}\\
& &    0.00 & 1747.7937 &$ -2.0419$&$[ -40,  40]$&$<  12.7$&$<13.91$&$ 13.70 \pm 0.05$\\
& &    0.00 & 1827.9351 &$ -1.6216$&$[ -20,  50]$&$  40.8 \pm   6.5$&$13.84 \pm 0.06$&\\
& &    0.00 & 2026.4768 &$ -0.9508$&$[   0,  50]$&$ 109.1 \pm  10.9$&$13.63 \pm 0.06$&\\
& &    0.00 & 2852.9642 &$  0.2577$&$[-100,  80]$&$ 597.8 \pm  11.3$&$>13.12$&\\
\ion{Mg}{2}\\
& &    0.00 & 2796.3520 &$ -0.2130$&$[-200, 100]$&$1927.7 \pm  13.4$&$>14.20$&$> 14.46$\\
& &    0.00 & 2803.5310 &$ -0.5151$&$[-200, 100]$&$1694.5 \pm  13.3$&$>14.46$&\\
\ion{Al}{2}\\
& &    0.00 & 1670.7874 &$  0.2742$&$[-180,  80]$&$ 718.0 \pm  11.8$&$>13.59$&$> 13.59$\\
\ion{Al}{3}\\
& &    0.00 & 1854.7164 &$ -0.2684$&$[ -50,  50]$&$ 122.2 \pm   6.7$&$12.97 \pm 0.03$&$ 12.95 \pm 0.02$\\
& &    0.00 & 1862.7895 &$ -0.5719$&$[ -50,  50]$&$  56.2 \pm   6.8$&$12.90 \pm 0.05$&\\
\ion{Si}{1}\\
& &    0.00 & 1845.5200 &$ -0.6402$&$[ -40,  40]$&$<  12.6$&$<12.44$&$< 12.44$\\
& &    0.00 & 2515.0730 &$ -0.7905$&$[ -40,  40]$&$<  17.8$&$<12.48$&\\
\ion{Si}{2}\\
&1/2 &    0.00 & 1526.7066 &$ -0.8962$&$[-190, 110]$&$ 659.0 \pm  18.6$&$>14.76$&$> 15.89$\\
&1/2 &    0.00 & 1808.0130 &$ -2.6603$&$[ -50,  50]$&$ 252.0 \pm   6.5$&$>15.89$&\\
&1/2 &    0.00 & 2335.1230 &$ -5.3716$&$[ -40,  40]$&$<  21.0$&$<17.19$&\\
&3/2 &  287.24 & 1533.4312 &$ -0.6396$&$[ -50,  30]$&$ 304.7 \pm   9.8$&$>14.26$&$> 14.26$\\
&3/2 &  287.24 & 1816.9285 &$ -2.7799$&$[ -40,  40]$&$<  12.7$&$<14.60$&\\
\ion{Si}{3}\\
& &    0.00 & 1892.0300 &$ -4.5702$&$[ -40,  40]$&$<  12.0$&$<16.33$&$< 16.33$\\
\ion{Ti}{2}\\
&1/2 &    0.00 & 1910.6000 &$ -0.6946$&$[ -40,  40]$&$<  11.5$&$<12.42$&$< 12.42$\\
\ion{Cr}{2}\\
& &    0.00 & 2056.2539 &$ -0.9788$&$[ -50,  50]$&$ 127.5 \pm  17.0$&$13.72 \pm 0.07$&$ 13.71 \pm 0.04$\\
& &    0.00 & 2062.2340 &$ -1.1079$&$[ -50,  50]$&$ 135.9 \pm  15.8$&$13.77 \pm 0.05$&\\
& &    0.00 & 2066.1610 &$ -1.2882$&$[ -40,  40]$&$<  30.4$&$13.53 \pm 0.13$&\\
\ion{Mn}{2}\\
& &    0.00 & 2576.8770 &$ -0.4549$&$[ -50,  50]$&$ 247.2 \pm   9.9$&$13.23 \pm 0.02$&$ 13.16 \pm 0.01$\\
& &    0.00 & 2594.4990 &$ -0.5670$&$[ -50,  50]$&$ 162.1 \pm   9.8$&$13.11 \pm 0.03$&\\
& &    0.00 & 2606.4620 &$ -0.7151$&$[ -50,  50]$&$ 133.9 \pm   9.1$&$13.14 \pm 0.03$&\\
\ion{Co}{2}\\
& &    0.00 & 1941.2852 &$ -1.4685$&$[ -40,  40]$&$<  12.9$&$<13.25$&$< 13.25$\\
\ion{Fe}{1}\\
& &    0.00 & 2484.0210 &$ -0.2541$&$[ -40,  40]$&$<  20.5$&$<12.02$&$< 12.02$\\
\ion{Fe}{2}\\
&9/2 &    0.00 & 1608.4511 &$ -1.2366$&$[-120,  70]$&$ 446.0 \pm  11.5$&$>14.92$&$ 15.22 \pm 0.03$\\
&9/2 &    0.00 & 1611.2005 &$ -2.8665$&$[ -50,  50]$&$  45.4 \pm   9.3$&$15.25 \pm 0.08$&\\
&9/2 &    0.00 & 2249.8768 &$ -2.7397$&$[ -50,  50]$&$ 110.8 \pm  13.0$&$15.23 \pm 0.05$&\\
&9/2 &    0.00 & 2260.7805 &$ -2.6126$&$[ -50,  50]$&$ 142.4 \pm  10.8$&$15.21 \pm 0.03$&\\
&9/2 &    0.00 & 2344.2140 &$ -0.9431$&$[-120,  80]$&$ 729.4 \pm  14.4$&$>14.57$&\\
&9/2 &    0.00 & 2374.4612 &$ -1.5045$&$[-100,  80]$&$ 546.6 \pm  13.3$&$>15.04$&\\
&9/2 &    0.00 & 2382.7650 &$ -0.4949$&$[-100,  80]$&$ 987.1 \pm  11.2$&$>14.27$&\\
&9/2 &    0.00 & 2586.6499 &$ -1.1605$&$[-130,  70]$&$ 779.0 \pm  13.5$&$>14.74$&\\
&9/2 &    0.00 & 2600.1729 &$ -0.6216$&$[-100, 100]$&$1075.3 \pm  10.7$&$>14.38$&\\
&7/2 &  384.79 & 2333.5160 &$ -1.1601$&$[ -40,  45]$&$ 184.8 \pm  10.1$&$13.89 \pm 0.03$&$ 13.88 \pm 0.02$\\
&7/2 &  384.79 & 2383.7884 &$ -2.2861$&$[ -40,  40]$&$<  19.9$&$<14.06$&\\
&7/2 &  384.79 & 2389.3582 &$ -1.0835$&$[ -50,  50]$&$ 232.1 \pm  10.4$&$13.90 \pm 0.02$&\\
&7/2 &  384.79 & 2396.3559 &$ -0.5414$&$[ -80,  80]$&$ 432.7 \pm  14.1$&$>13.85$&\\
&7/2 &  384.79 & 2626.4511 &$ -1.3556$&$[ -40,  50]$&$ 143.5 \pm   9.7$&$13.84 \pm 0.03$&\\
&5/2 &  667.68 & 2328.1110 &$ -1.4492$&$[ -40,  50]$&$  58.2 \pm  12.0$&$13.61 \pm 0.08$&$ 13.63 \pm 0.01$\\
&5/2 &  667.68 & 2349.0220 &$ -1.0841$&$[ -60,  60]$&$ 211.1 \pm  11.8$&$13.86 \pm 0.03$&\\
&5/2 &  667.68 & 2399.9728 &$ -0.9255$&$[ -40,  50]$&$ 198.7 \pm  11.1$&$13.67 \pm 0.03$&\\
&5/2 &  667.68 & 2607.8664 &$ -0.9281$&$[ -40,  40]$&$ 193.3 \pm   8.2$&$13.57 \pm 0.02$&\\
&5/2 &  667.68 & 2618.3991 &$ -1.2967$&$[ -40,  40]$&$  95.7 \pm   8.7$&$13.57 \pm 0.04$&\\
&3/2 &  862.61 & 2338.7250 &$ -1.0339$&$[ -30,  40]$&$  69.2 \pm   9.1$&$13.26 \pm 0.06$&$ 13.33 \pm 0.02$\\
&3/2 &  862.61 & 2359.8280 &$ -1.2421$&$[ -40,  40]$&$  65.7 \pm  11.5$&$13.46 \pm 0.07$&\\
&3/2 &  862.61 & 2411.2533 &$ -0.6778$&$[ -40,  40]$&$ 173.1 \pm   9.9$&$13.34 \pm 0.03$&\\
&3/2 &  862.61 & 2621.1912 &$ -2.4067$&$[ -40,  40]$&$<  18.7$&$<14.08$&\\
&1/2 &  977.05 & 2345.0010 &$ -0.8027$&$[ -25,  40]$&$  70.8 \pm   8.8$&$13.05 \pm 0.05$&$ 13.05 \pm 0.03$\\
&1/2 &  977.05 & 2414.0450 &$ -0.7557$&$[ -40,  40]$&$ 305.7 \pm   8.4$&$>13.93$&\\
&1/2 &  977.05 & 2622.4518 &$ -1.2518$&$[ -40,  40]$&$  28.5 \pm   9.3$&$12.98 \pm 0.13$&\\
&1/2 &  977.05 & 2629.0777 &$ -0.7620$&$[ -40,  50]$&$ 104.7 \pm   9.9$&$13.07 \pm 0.04$&\\
\ion{Ni}{2}\\
& &    0.00 & 1703.4050 &$ -2.2218$&$[ -40,  40]$&$<  13.1$&$<14.12$&$ 13.84 \pm 0.03$\\
& &    0.00 & 1709.6042 &$ -1.4895$&$[ -40,  40]$&$  50.4 \pm   6.6$&$13.83 \pm 0.06$&\\
& &    0.00 & 1741.5531 &$ -1.3696$&$[ -50,  50]$&$  70.8 \pm   7.4$&$13.85 \pm 0.05$&\\
& &    0.00 & 1751.9157 &$ -1.5575$&$[ -40,  40]$&$  43.9 \pm   6.6$&$13.81 \pm 0.06$&\\
\ion{Zn}{2}\\
& &    0.00 & 2026.1360 &$ -0.3107$&$[ -50,  50]$&$ 147.7 \pm  14.9$&$13.03 \pm 0.05$&$ 13.02 \pm 0.04$\\
& &    0.00 & 2062.6640 &$ -0.5918$&$[ -40,  25]$&$  79.2 \pm  12.8$&$13.00 \pm 0.07$&\\
\enddata
\tablenotetext{a}{Total angular momentum of the electron spin and orbital angular moment.
$E_{J}$ is the energy above the ground state.}
\tablenotetext{b}{Velocity interval over which the equivalent width and
column density are measured.}
\tablenotetext{c}{Rest equivalent width.}
\end{deluxetable}

\begin{deluxetable}{lcccc}
\tablewidth{0pc}
\tablecaption{SUMMARY TABLE FOR THE ISM IN GRB HOST GALAXIES\label{tab:summ}}
\tabletypesize{\footnotesize}
\tablehead{& 
\colhead{HG~050730} &
\colhead{HG~050820} &
\colhead{HG~051111} &
\colhead{HG~060418} }
\startdata
$z$ & 3.9686& 2.6147& 1.5495& 1.4900\\
$\log \mnhi$ &$22.15^{+0.10}_{-0.10}$&$21.00^{+0.10}_{-0.10}$&$23.00$&$23.00$\\
$\lbrack$M/H]$^a$&$ -2.26$&$ -0.63$&$> -2.78$&$> -2.65$\\
$\log \N{Fe^+}^b$&15.15&14.90&15.36&15.26\\
$\lbrack$C/Fe] &$>-1.04$&$>-0.45$&$...$&$...$\\
$\lbrack$N/Fe] &$-0.85$&$<+ 2.89$&$...$&$...$\\
$\lbrack$O/Fe] &$>-0.89$&$>-0.30$&$...$&$...$\\
$\lbrack$Mg/Fe] &$<+ 0.85$&$+ 0.88$&$>-0.83$&$>-0.88$\\
$\lbrack$Al/Fe] &$>-0.60$&$>+ 0.20$&$>-0.61$&$>-0.66$\\
$\lbrack$Si/Fe] &$>-0.28$&$>+ 0.48$&$>+ 0.75$&$>+ 0.58$\\
$\lbrack$P/Fe] &$<+ 1.04$&$+ 0.71$&$...$&$...$\\
$\lbrack$S/Fe] &$+ 0.24$&$+ 0.97$&$...$&$...$\\
$\lbrack$Ti/Fe] &$...$&$...$&$<-0.39$&$<-0.27$\\
$\lbrack$Cr/Fe] &$...$&$+ 0.26$&$+ 0.35$&$+ 0.28$\\
$\lbrack$Mn/Fe] &$...$&$...$&$>+ 0.25$&$-0.12$\\
$\lbrack$Co/Fe] &$...$&$...$&$<+ 0.38$&$<+ 0.58$\\
$\lbrack$Ni/Fe] &$-0.22$&$+ 0.04$&$-0.15$&$-0.17$\\
$\lbrack$Zn/Fe] &$...$&$+ 0.89$&$>+ 1.18$&$+ 0.59$\\
$\log \N{O^0}_{J=1}$&$> 14.94$&$$&$$&$$\\
$\log \N{O^0}_{J=0}$&$> 14.61$&$< 13.41$&$$&$$\\
$\log \N{Si^+}_{J=3/2}$&$> 14.56$&$ 13.73 \pm 0.02$&$ 15.00 \pm 0.06$&$> 14.26$\\
$\log \N{Fe^+}_{J=7/2}$&$ 14.29 \pm 0.04$&$< 13.31$&$ 14.01 \pm 0.01$&$ 13.88 \pm 0.02$\\
$\log \N{Fe^+}_{J=5/2}$&$ 14.12 \pm 0.07$&$< 13.29$&$ 13.78 \pm 0.01$&$ 13.63 \pm 0.01$\\
$\log \N{Fe^+}_{J=3/2}$&$ 13.92 \pm 0.06$&$< 13.15$&$ 13.59 \pm 0.01$&$ 13.33 \pm 0.02$\\
$\log \N{Fe^+}_{J=1/2}$&$ 13.65 \pm 0.07$&$< 12.97$&$ 13.26 \pm 0.01$&$ 13.05 \pm 0.03$\\
$\log \N{Mg^0}$&$$&$< 12.60$&$ 14.68 \pm 0.04$&$ 13.70 \pm 0.05$\\
\enddata
\tablenotetext{a}{Metallicity based on S, Si, and Zn in that order of preference.
Lower limits assume $\mnhi = 10^{23} \cm{-2}$.}
\tablenotetext{b}{Corrected for the population of the excited levels
for this and all of the following entries.}
\tablecomments{We assume the solar abundance measurements given by
\cite{solabnd} throughout this Table.  Furthermore, we do not adopt
ionization corrections or differential depletion corrections.}
\end{deluxetable}

\begin{deluxetable}{lcccccc}
\tablewidth{0pc}
\tablecaption{IONIC COLUMN DENSITIES FOR GRB~050730 MgII $z$=1.773\label{tab:050730MgII1}}
\tabletypesize{\footnotesize}
\tablehead{\colhead{Ion} & \colhead{$\lambda$} & \colhead{$\log f$}
& \colhead{$v_{int}^a$} 
& \colhead{$W_\lambda^b$} 
& \colhead{$\log N$} & \colhead{$\log N_{adopt}$} \\
& (\AA) & & (\kms) & (m\AA) & & }
\startdata
\ion{Mg}{1}\\
&2852.9642 &$  0.2577$&$[ -80,  90]$&$ 115.1 \pm  34.4$&$ 12.14 \pm 0.10$&$ 12.15 \pm 0.10$\\
\ion{Mg}{2}\\
&2796.3520 &$ -0.2130$&$[ -80,  90]$&$ 943.6 \pm  22.7$&$> 13.79$&$> 14.02$\\
&2803.5310 &$ -0.5151$&$[ -80,  90]$&$ 797.5 \pm  21.7$&$> 14.02$& \\
\ion{Fe}{2}\\
&2260.7805 &$ -2.6126$&$[ -80,  90]$&$<  53.9$&$< 14.88$&$ 14.02 \pm 0.03$\\
&2344.2140 &$ -0.9431$&$[ -80,  90]$&$ 264.8 \pm  22.5$&$ 13.93 \pm 0.06$& \\
&2374.4612 &$ -1.5045$&$[ -80,  90]$&$ 311.6 \pm  20.6$&$< 14.42$& \\
&2382.7650 &$ -0.4949$&$[ -80,  90]$&$ 398.7 \pm  19.1$&$> 13.80$& \\
&2586.6500 &$ -1.1605$&$[ -80,  90]$&$ 312.9 \pm  19.0$&$ 14.06 \pm 0.03$& \\
&2600.1729 &$ -0.6216$&$[ -80,  90]$&$ 436.9 \pm  23.0$&$> 13.85$& \\
\enddata
\tablenotetext{a}{Velocity interval over which the equivalent width and
column density are measured.}
\tablenotetext{b}{Rest equivalent width.}
\end{deluxetable}

\begin{deluxetable}{lcccccc}
\tablewidth{0pc}
\tablecaption{IONIC COLUMN DENSITIES FOR GRB~050730 MgII $z$=2.253\label{tab:050730MgII2}}
\tabletypesize{\footnotesize}
\tablehead{\colhead{Ion} & \colhead{$\lambda$} & \colhead{$\log f$}
& \colhead{$v_{int}^a$} 
& \colhead{$W_\lambda^b$} 
& \colhead{$\log N$} & \colhead{$\log N_{adopt}$} \\
& (\AA) & & (\kms) & (m\AA) & & }
\startdata
\ion{Mg}{1}\\
&2852.9642 &$  0.2577$&$[ -60, 120]$&$ 471.6 \pm  33.0$&$ 12.71 \pm 0.04$&$ 12.71 \pm 0.04$\\
\ion{Mg}{2}\\
&2796.3520 &$ -0.2130$&$[ -60, 120]$&$ 886.4 \pm  31.1$&$> 13.74$&$> 13.97$\\
&2803.5310 &$ -0.5151$&$[ -60, 130]$&$ 825.7 \pm  38.4$&$> 13.97$& \\
\ion{Fe}{2}\\
&2260.7805 &$ -2.6126$&$[ -50, 120]$&$<  41.8$&$< 14.76$&$> 14.55$\\
&2374.4612 &$ -1.5045$&$[ -50, 120]$&$ 255.1 \pm  27.2$&$> 14.55$& \\
&2382.7650 &$ -0.4949$&$[ -50, 120]$&$ 738.2 \pm  26.5$&$> 14.11$& \\
&2586.6500 &$ -1.1605$&$[ -50, 120]$&$ 282.0 \pm  25.2$&$> 14.38$& \\
&2600.1729 &$ -0.6216$&$[ -20, 120]$&$ 424.7 \pm  23.7$&$> 13.87$& \\
\enddata
\tablenotetext{a}{Velocity interval over which the equivalent width and
column density are measured.}
\tablenotetext{b}{Rest equivalent width.}
\end{deluxetable}

\begin{deluxetable}{lcccccc}
\tablewidth{0pc}
\tablecaption{IONIC COLUMN DENSITIES FOR GRB~050820 MgII $z$=0.692\label{tab:050820MgII1}}
\tabletypesize{\footnotesize}
\tablehead{\colhead{Ion} & \colhead{$\lambda$} & \colhead{$\log f$}
& \colhead{$v_{int}^a$} 
& \colhead{$W_\lambda^b$} 
& \colhead{$\log N$} & \colhead{$\log N_{adopt}$} \\
& (\AA) & & (\kms) & (m\AA) & & }
\startdata
\ion{Mg}{2}\\
&2796.3520 &$ -0.2130$&$[-480, 180]$&$2991.9 \pm  27.5$&$> 14.28$&$> 14.40$\\
&2803.5310 &$ -0.5151$&$[-480, 180]$&$2343.0 \pm  30.5$&$> 14.40$& \\
\ion{Ca}{2}\\
&3934.7770 &$ -0.1871$&$[-200,   0]$&$ 210.3 \pm  22.1$&$ 12.49 \pm 0.04$&$ 12.52 \pm 0.03$\\
&3969.5910 &$ -0.4921$&$[-200,   0]$&$ 140.0 \pm  16.5$&$ 12.56 \pm 0.05$& \\
\ion{Fe}{2}\\
&2344.2140 &$ -0.9431$&$[-450, 150]$&$1229.6 \pm  52.6$&$> 14.65$&$> 14.65$\\
&2382.7650 &$ -0.4949$&$[-480, 180]$&$1722.5 \pm  47.3$&$> 14.31$& \\
\enddata
\tablenotetext{a}{Velocity interval over which the equivalent width and
column density are measured.}
\tablenotetext{b}{Rest equivalent width.}
\end{deluxetable}

\begin{deluxetable}{lcccccc}
\tablewidth{0pc}
\tablecaption{IONIC COLUMN DENSITIES FOR GRB~050820 MgII $z$=1.430\label{tab:050820MgII2}}
\tabletypesize{\footnotesize}
\tablehead{\colhead{Ion} & \colhead{$\lambda$} & \colhead{$\log f$}
& \colhead{$v_{int}^a$} 
& \colhead{$W_\lambda^b$} 
& \colhead{$\log N$} & \colhead{$\log N_{adopt}$} \\
& (\AA) & & (\kms) & (m\AA) & & }
\startdata
\ion{Mg}{1}\\
&2852.9642 &$  0.2577$&$[-200, 150]$&$ 935.1 \pm 105.5$&$ 11.74 \pm 0.11$&$ 11.74 \pm 0.11$\\
\ion{Mg}{2}\\
&2796.3520 &$ -0.2130$&$[ -80, 180]$&$1935.7 \pm 134.2$&$> 13.45$&$> 14.20$\\
&2803.5310 &$ -0.5151$&$[-200, 180]$&$1084.2 \pm  20.0$&$> 14.20$& \\
\ion{Fe}{2}\\
&2260.7805 &$ -2.6126$&$[-200, 100]$&$<  27.7$&$< 14.59$&$ 14.29 \pm 0.03$\\
&2344.2140 &$ -0.9431$&$[-200, 150]$&$ 365.4 \pm  18.4$&$> 14.24$& \\
&2374.4612 &$ -1.5045$&$[-200, 150]$&$ 193.0 \pm  14.6$&$ 14.29 \pm 0.03$& \\
&2382.7650 &$ -0.4949$&$[-200, 150]$&$ 557.4 \pm  17.0$&$> 13.95$& \\
&2586.6500 &$ -1.1605$&$[-200, 150]$&$ 391.1 \pm  15.8$&$> 14.33$& \\
&2600.1729 &$ -0.6216$&$[-200, 150]$&$ 533.4 \pm  17.2$&$> 14.03$& \\
\enddata
\tablenotetext{a}{Velocity interval over which the equivalent width and
column density are measured.}
\tablenotetext{b}{Rest equivalent width.}
\end{deluxetable}

\begin{deluxetable}{lcccccc}
\tablewidth{0pc}
\tablecaption{IONIC COLUMN DENSITIES FOR GRB~051111 MgII $z$=1.189\label{tab:051111MgII1}}
\tabletypesize{\footnotesize}
\tablehead{\colhead{Ion} & \colhead{$\lambda$} & \colhead{$\log f$}
& \colhead{$v_{int}^a$} 
& \colhead{$W_\lambda^b$} 
& \colhead{$\log N$} & \colhead{$\log N_{adopt}$} \\
& (\AA) & & (\kms) & (m\AA) & & }
\startdata
\ion{Mg}{1}\\
&2852.9652 &$  0.2577$&$[-130, 180]$&$ 490.7 \pm  10.2$&$ 12.69 \pm 0.01$&$ 12.69 \pm 0.01$\\
\ion{Mg}{2}\\
&2796.3520 &$ -0.2130$&$[-130, 180]$&$2027.2 \pm  10.0$&$> 14.27$&$> 14.49$\\
&2803.5310 &$ -0.5151$&$[-130, 180]$&$1719.2 \pm   9.1$&$> 14.49$& \\
\ion{Ti}{2}\\
&3242.9290 &$ -0.6345$&$[   0, 150]$&$<  17.4$&$ 12.13 \pm 0.13$&$ 12.14 \pm 0.13$\\
\ion{Fe}{2}\\
&2260.7805 &$ -2.6126$&$[-130, 180]$&$  49.7 \pm  12.2$&$< 14.76$&$ 14.45 \pm 0.02$\\
&2344.2140 &$ -0.9431$&$[-130, 180]$&$ 817.3 \pm  10.9$&$> 14.48$& \\
&2374.4612 &$ -1.5045$&$[-130, 180]$&$ 316.5 \pm  13.4$&$ 14.45 \pm 0.02$& \\
&2382.7650 &$ -0.4949$&$[-130, 180]$&$1184.8 \pm  11.2$&$> 14.33$& \\
&2600.1729 &$ -0.6216$&$[-130, 180]$&$1288.0 \pm  13.7$&$> 14.39$& \\
\ion{Zn}{2}\\
&2026.1360 &$ -0.3107$&$[   0, 150]$&$  48.7 \pm  11.1$&$< 12.56$&$< 12.56$\\
\enddata
\tablenotetext{a}{Velocity interval over which the equivalent width and
column density are measured.}
\tablenotetext{b}{Rest equivalent width.}
\end{deluxetable}

\begin{deluxetable}{lcccccc}
\tablewidth{0pc}
\tablecaption{IONIC COLUMN DENSITIES FOR GRB~060418 MgII $z$=0.603\label{tab:060418MgII1}}
\tabletypesize{\footnotesize}
\tablehead{\colhead{Ion} & \colhead{$\lambda$} & \colhead{$\log f$}
& \colhead{$v_{int}^a$} 
& \colhead{$W_\lambda^b$} 
& \colhead{$\log N$} & \colhead{$\log N_{adopt}$} \\
& (\AA) & & (\kms) & (m\AA) & & }
\startdata
\ion{Mg}{1}\\
&2852.9652 &$  0.2577$&$[-180,  30]$&$ 383.3 \pm  14.4$&$ 12.58 \pm 0.02$&$ 12.58 \pm 0.02$\\
\ion{Mg}{2}\\
&2796.3520 &$ -0.2130$&$[-180,  30]$&$1270.4 \pm  13.4$&$> 14.03$&$> 14.29$\\
&2803.5310 &$ -0.5151$&$[-180,  30]$&$1214.7 \pm  13.9$&$> 14.29$& \\
\ion{Mn}{2}\\
&2576.8770 &$ -0.4549$&$[-100,  30]$&$ 140.2 \pm  15.3$&$ 12.89 \pm 0.05$&$ 12.89 \pm 0.05$\\
\ion{Fe}{2}\\
&2260.7805 &$ -2.6126$&$[-180,  30]$&$ 311.1 \pm  40.6$&$> 15.67$&$> 15.67$\\
&2344.2140 &$ -0.9431$&$[-180,  30]$&$ 711.9 \pm  27.4$&$> 14.53$& \\
&2374.4612 &$ -1.5045$&$[-180,  30]$&$ 579.3 \pm  26.4$&$> 14.96$& \\
&2382.7650 &$ -0.4949$&$[-180,  30]$&$ 943.3 \pm  25.6$&$> 14.23$& \\
&2586.6499 &$ -1.1605$&$[-180,  30]$&$ 737.8 \pm  15.9$&$> 14.73$& \\
&2600.1729 &$ -0.6216$&$[-180,  30]$&$ 938.6 \pm  16.9$&$> 14.31$& \\
\enddata
\tablenotetext{a}{Velocity interval over which the equivalent width and
column density are measured.}
\tablenotetext{b}{Rest equivalent width.}
\end{deluxetable}

\begin{deluxetable}{lcccccc}
\tablewidth{0pc}
\tablecaption{IONIC COLUMN DENSITIES FOR GRB~060418 MgII $z$=0.656\label{tab:060418MgII2}}
\tabletypesize{\footnotesize}
\tablehead{\colhead{Ion} & \colhead{$\lambda$} & \colhead{$\log f$}
& \colhead{$v_{int}^a$} 
& \colhead{$W_\lambda^b$} 
& \colhead{$\log N$} & \colhead{$\log N_{adopt}$} \\
& (\AA) & & (\kms) & (m\AA) & & }
\startdata
\ion{Mg}{1}\\
&2852.9652 &$  0.2577$&$[-100,  60]$&$<  26.8$&$< 11.50$&$< 11.50$\\
\ion{Mg}{2}\\
&2796.3520 &$ -0.2130$&$[-100,  60]$&$ 972.5 \pm   9.9$&$> 13.81$&$> 13.95$\\
&2803.5310 &$ -0.5151$&$[-100,  60]$&$ 793.9 \pm  10.6$&$> 13.95$& \\
\ion{Fe}{2}\\
&2260.7805 &$ -2.6126$&$[-100,  60]$&$<  56.8$&$< 14.92$&$ 13.86 \pm 0.02$\\
&2344.2140 &$ -0.9431$&$[-100,  60]$&$ 298.7 \pm  19.7$&$ 13.92 \pm 0.04$& \\
&2374.4612 &$ -1.5045$&$[-100,  60]$&$ 107.9 \pm  18.9$&$ 13.99 \pm 0.07$& \\
&2382.7650 &$ -0.4949$&$[-100,  60]$&$ 460.0 \pm  17.7$&$> 13.76$& \\
&2586.6499 &$ -1.1605$&$[-100,  60]$&$ 196.7 \pm  14.2$&$ 13.81 \pm 0.03$& \\
&2600.1729 &$ -0.6216$&$[-100,  60]$&$ 491.1 \pm  12.1$&$> 13.84$& \\
\enddata
\tablenotetext{a}{Velocity interval over which the equivalent width and
column density are measured.}
\tablenotetext{b}{Rest equivalent width.}
\end{deluxetable}

\begin{deluxetable}{lccccccccc}
\tablewidth{0pc}
\tablecaption{IONIC COLUMN DENSITIES FOR GRB~060418 MgII $z$=1.107\label{tab:060418MgII3}}
\tabletypesize{\footnotesize}
\tablehead{\colhead{Ion} & \colhead{$J^a$} & \colhead{$E_{J}$} & \colhead{$\lambda$} & \colhead{$\log f$}
& \colhead{$v_{int}^b$} 
& \colhead{$W_\lambda^c$} & \colhead{$\log N$} & \colhead{$\log N_{adopt}$} \\
& & (cm$^{-1}$) & (\AA) & & (\kms) & (m\AA) }
\startdata
\ion{Mg}{1}\\
& &    0.00 & 2852.9642 &$  0.2577$&$[ -80, 100]$&$ 469.5 \pm  16.7$&$>12.90$&$> 12.90$\\
\ion{Mg}{2}\\
& &    0.00 & 2796.3520 &$ -0.2130$&$[-120, 200]$&$1841.4 \pm  22.6$&$>14.09$&$> 14.32$\\
& &    0.00 & 2803.5310 &$ -0.5151$&$[-120, 200]$&$1536.1 \pm  20.1$&$>14.32$&\\
\ion{Al}{2}\\
& &    0.00 & 1670.7874 &$  0.2742$&$[-100, 200]$&$ 738.0 \pm  42.4$&$>13.52$&$> 13.52$\\
\ion{Al}{3}\\
& &    0.00 & 1854.7164 &$ -0.2684$&$[ -50, 100]$&$ 198.5 \pm  14.2$&$13.22 \pm 0.04$&$ 13.20 \pm 0.03$\\
& &    0.00 & 1862.7895 &$ -0.5719$&$[ -50, 100]$&$  96.0 \pm  14.4$&$13.16 \pm 0.06$&\\
\ion{Si}{2}\\
&1/2 &    0.00 & 1808.0130 &$ -2.6603$&$[ -70,  50]$&$ 123.8 \pm  16.2$&$15.43 \pm 0.06$&$ 15.43 \pm 0.06$\\
\ion{Ca}{2}\\
& &    0.00 & 3934.7770 &$ -0.1871$&$[ -50,  50]$&$ 144.4 \pm  12.9$&$12.34 \pm 0.03$&$ 12.34 \pm 0.04$\\
\ion{Ti}{2}\\
&1/2 &    0.00 & 3242.9290 &$ -0.6345$&$[ -50,  50]$&$<  23.4$&$<12.23$&$< 11.97$\\
&1/2 &    0.00 & 3384.7400 &$ -0.4461$&$[ -50,  50]$&$<  21.9$&$<11.97$&\\
\ion{Cr}{2}\\
& &    0.00 & 2056.2539 &$ -0.9788$&$[ -50,  50]$&$<  18.4$&$<12.86$&$< 12.86$\\
\ion{Mn}{2}\\
& &    0.00 & 2576.8770 &$ -0.4549$&$[ -50,  50]$&$ 121.1 \pm  20.2$&$12.87 \pm 0.08$&$ 12.82 \pm 0.06$\\
& &    0.00 & 2594.4990 &$ -0.5670$&$[ -50,  50]$&$  76.3 \pm  17.0$&$12.78 \pm 0.09$&\\
\ion{Fe}{2}\\
&9/2 &    0.00 & 2260.7805 &$ -2.6126$&$[ -50,  50]$&$  45.7 \pm   7.9$&$14.66 \pm 0.07$&$ 14.66 \pm 0.02$\\
&9/2 &    0.00 & 2344.2140 &$ -0.9431$&$[ -80, 150]$&$ 769.3 \pm  12.1$&$>14.56$&\\
&9/2 &    0.00 & 2374.4612 &$ -1.5045$&$[ -80, 150]$&$ 433.1 \pm  18.0$&$14.66 \pm 0.03$&\\
&9/2 &    0.00 & 2382.7650 &$ -0.4949$&$[ -80, 150]$&$1005.6 \pm  15.5$&$>14.16$&\\
&9/2 &    0.00 & 2600.1729 &$ -0.6216$&$[ -80, 150]$&$1104.6 \pm  25.1$&$>14.27$&\\
\ion{Ni}{2}\\
& &    0.00 & 1741.5531 &$ -1.3696$&$[ -50,  50]$&$<  35.1$&$<13.75$&$< 13.75$\\
\ion{Zn}{2}\\
& &    0.00 & 2026.1360 &$ -0.3107$&$[ -50,  50]$&$ 118.6 \pm   8.9$&$12.98 \pm 0.03$&$ 12.96 \pm 0.03$\\
& &    0.00 & 2062.6640 &$ -0.5918$&$[ -50,  40]$&$  70.0 \pm   7.7$&$12.95 \pm 0.05$&\\
\enddata
\tablenotetext{a}{Total angular momentum of the electron spin and orbital angular moment.
$E_{J}$ is the energy above the ground state.}
\tablenotetext{b}{Velocity interval over which the equivalent width and
column density are measured.}
\tablenotetext{c}{Rest equivalent width.}
\end{deluxetable}

\begin{deluxetable}{lccccccc}
\tablewidth{0pc}
\tablecaption{\ion{Mg}{2} SUMMARY TABLE\label{tab:mgsumm}}
\tabletypesize{\footnotesize}
\tablehead{\colhead{GRB} & \colhead{$z$} & \colhead{$W_\lambda$}
& \colhead{$\N{Fe^+}$} & \colhead{$\N{Zn^+}$} 
& \colhead{$\N{Mg^0}$} 
& \colhead{$\mnhi^a$} & \colhead{$D^b$} \\
& & (\AA) & & & & }
\startdata
050730 &1.773 & 943.6&$ 14.02 \pm 0.03$&$$&$ 12.15 \pm 0.10$&18.5&7.8\\
050730 &2.253 & 886.4&$> 14.55$&$$&$ 12.71 \pm 0.04$&19.0&7.2\\
050820 &0.692 &2991.9&$> 14.65$&$$&$$&19.1&7.8\\
050820 &1.430 &1935.7&$ 14.29 \pm 0.03$&$$&$ 11.74 \pm 0.11$&18.8&8.1\\
051111 &1.189 &2027.2&$ 14.45 \pm 0.02$&$< 12.56$&$ 12.69 \pm 0.01$&19.0&8.5\\
060418 &0.603 &1270.4&$> 15.67$&$$&$ 12.58 \pm 0.02$&20.2&8.2\\
060418 &0.656 & 972.5&$ 13.86 \pm 0.02$&$$&$< 11.50$&18.4&7.2\\
060418 &1.107 &1841.4&$ 14.66 \pm 0.02$&$ 12.96 \pm 0.03$&$> 12.90$&20.3&7.5\\
\enddata
\tablenotetext{a}{Minimum \nhi\ value based on the measured 
 Fe$^+$ and/or Zn$^+$ column densities, assuming the metallcity
is sub-solar.}
\tablenotetext{b}{D-statistic \citep{ellison06}, where 
values larger than 6.8 indicate a greater likelihood that the 
absorber has $\log \mnhi > 20.3$.}
\end{deluxetable}

\clearpage

\begin{figure}
\epsscale{0.85}
\plotone{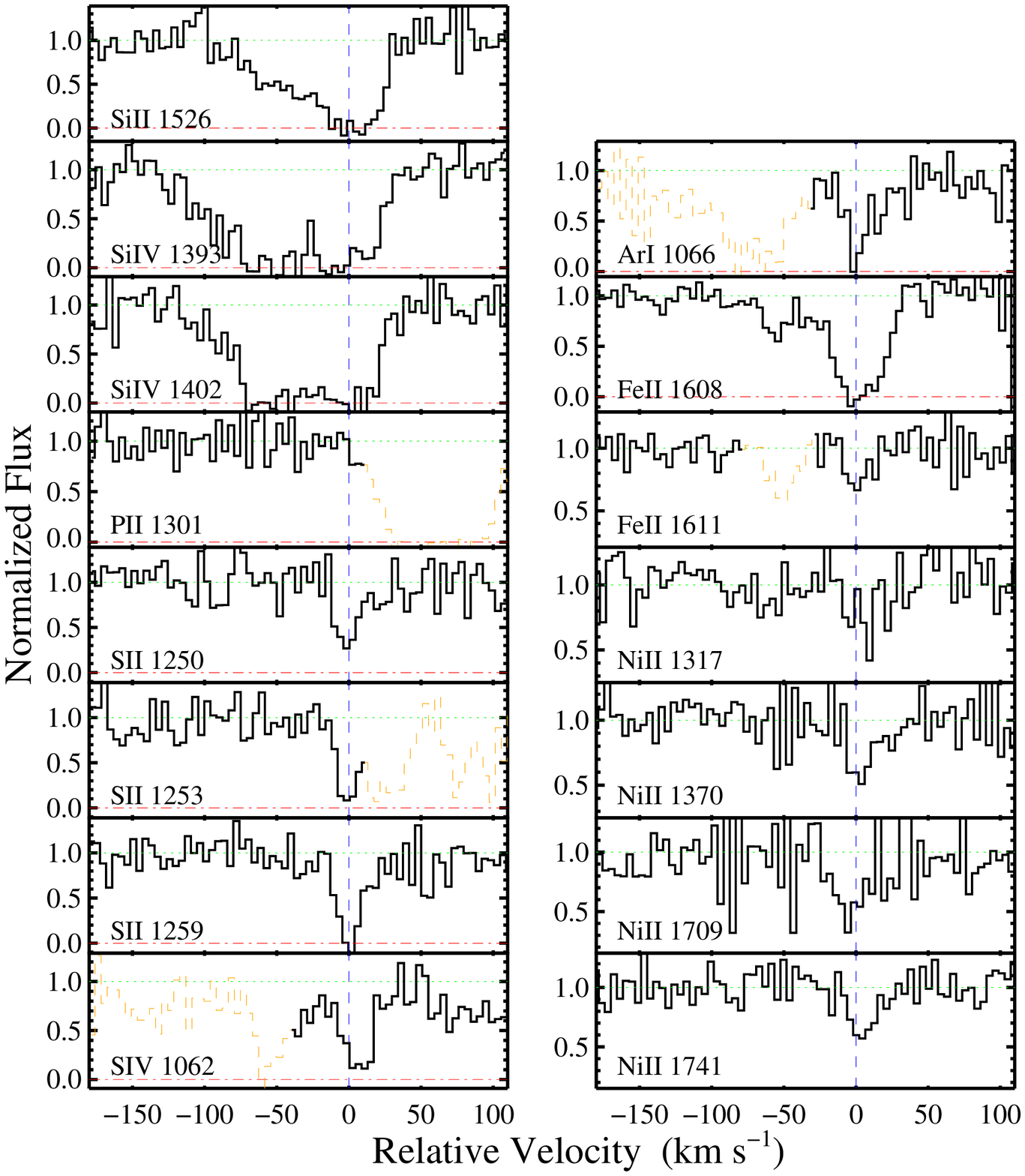}
\end{figure}


\begin{figure}
\epsscale{0.85}
\plotone{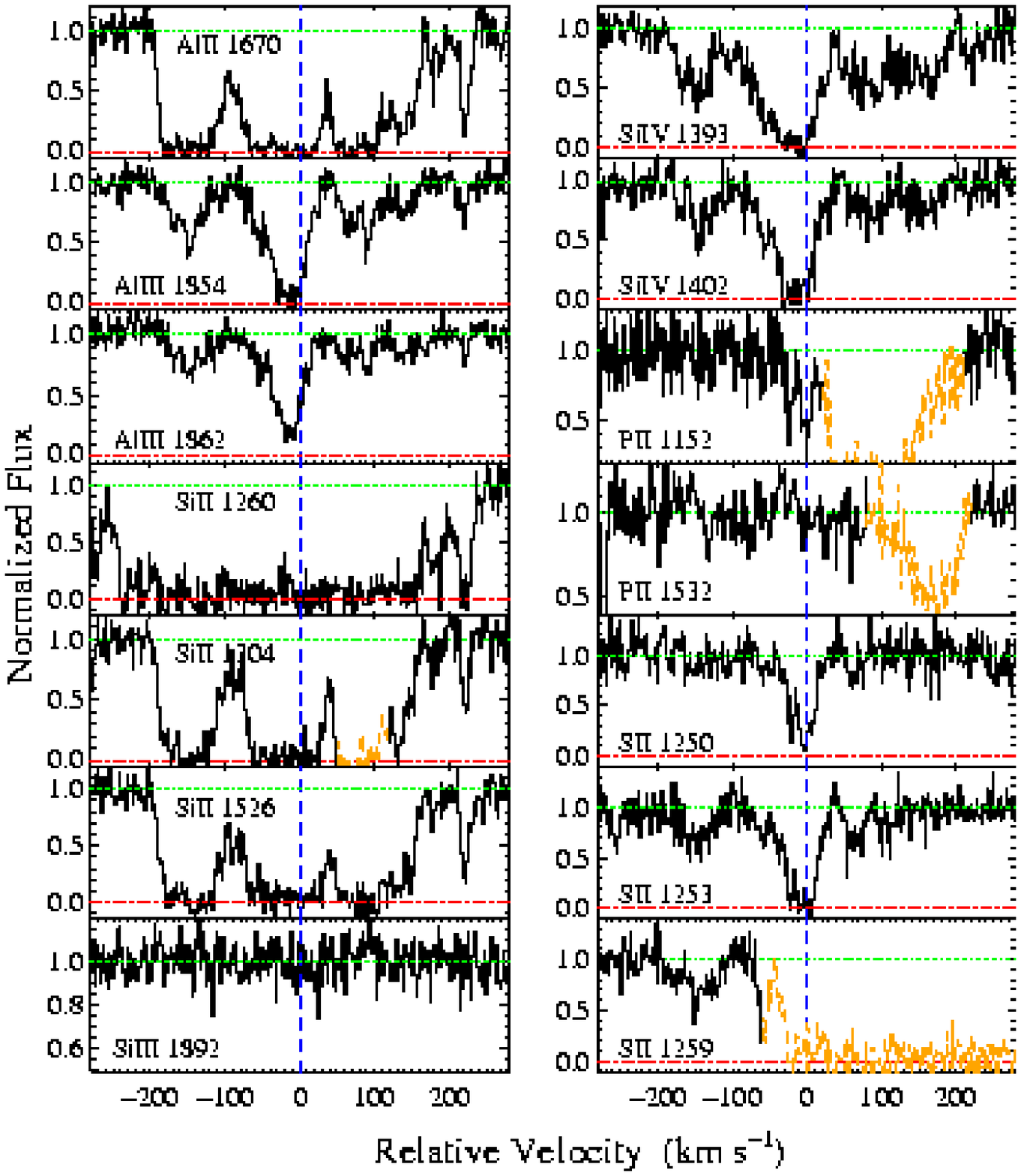}
\end{figure}

\begin{figure}
\epsscale{0.85}
\plotone{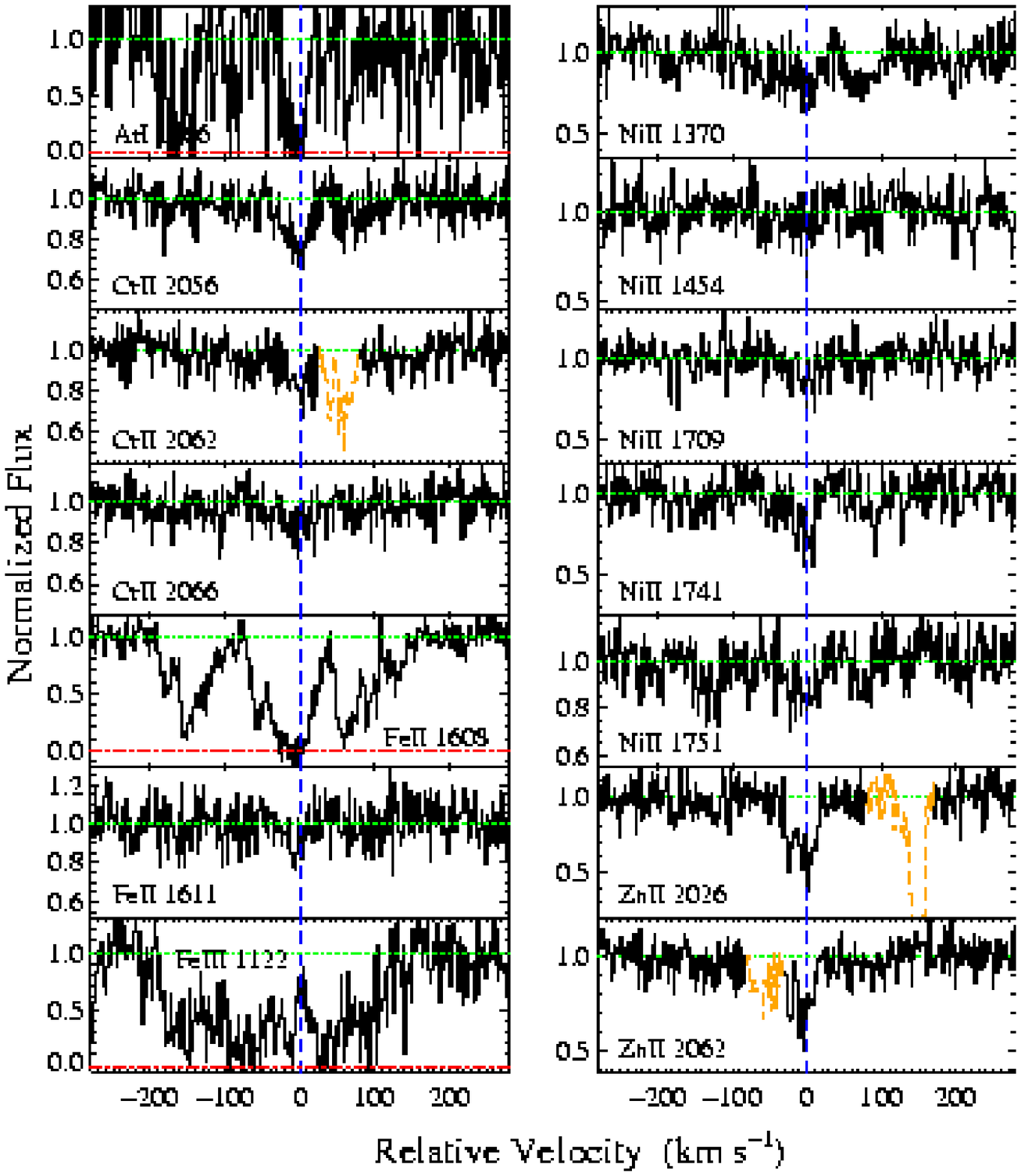}
\end{figure}

\clearpage

\begin{figure}
\epsscale{0.85}
\plotone{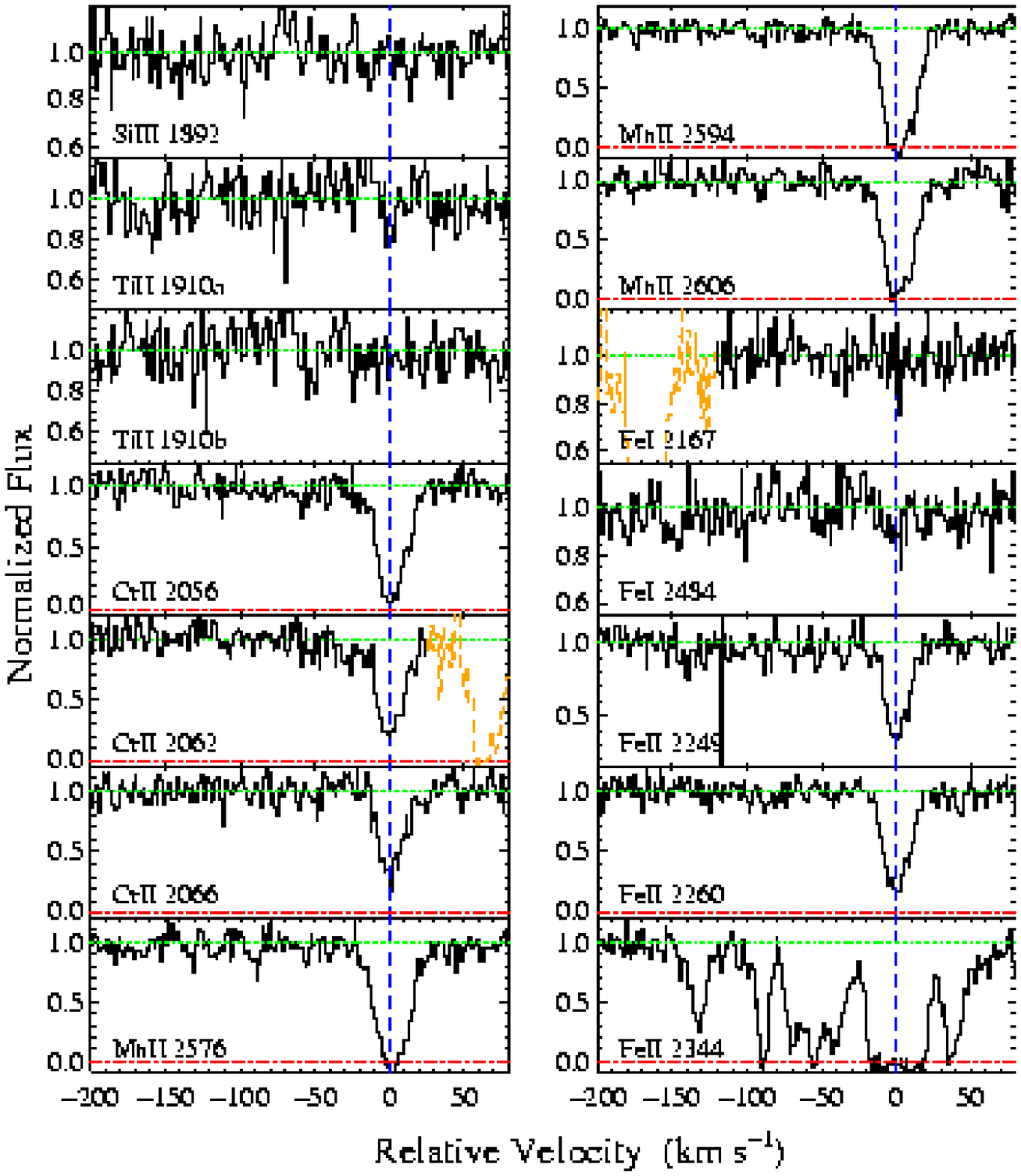}
\end{figure}

\begin{figure}
\epsscale{0.85}
\plotone{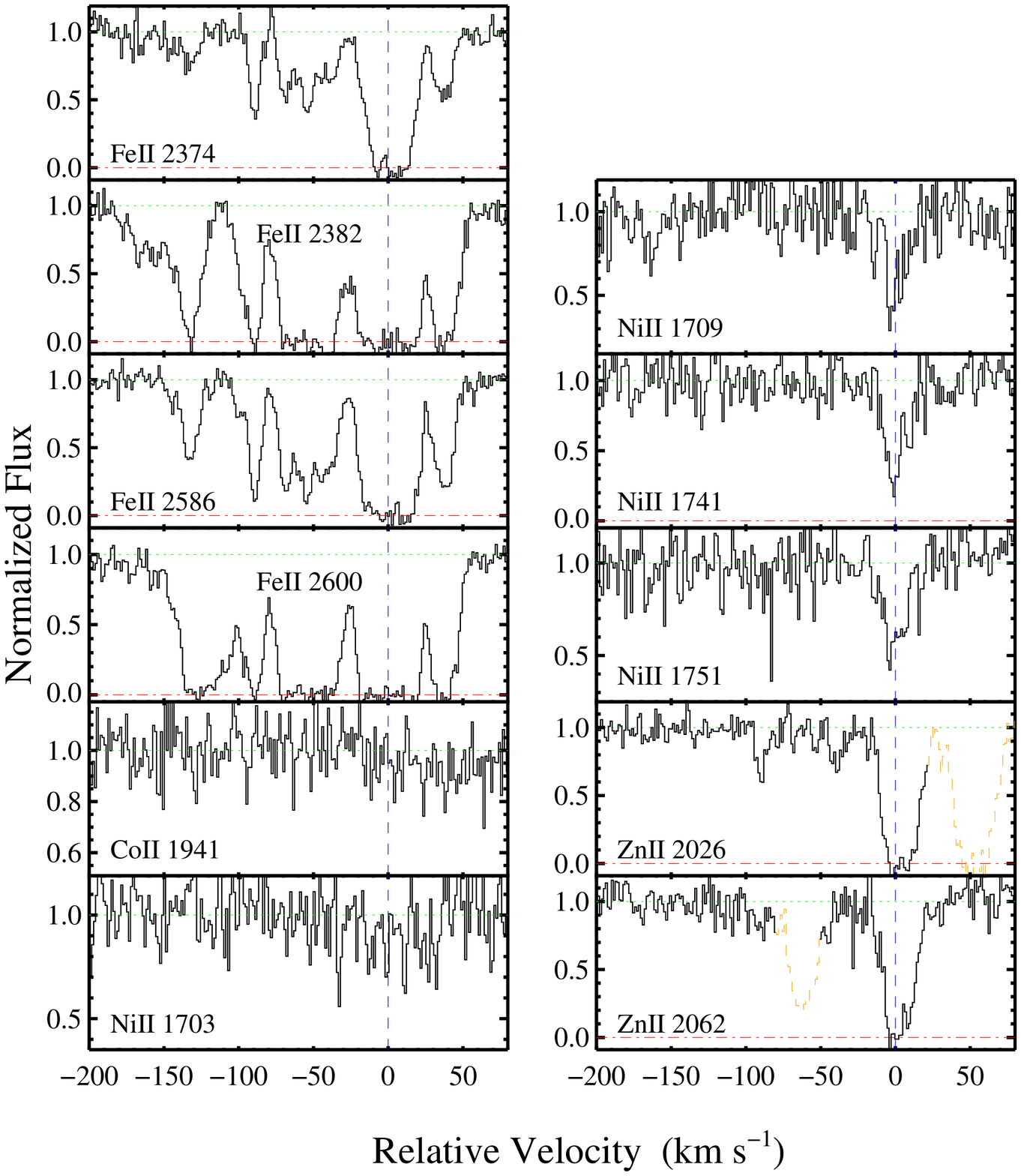}
\end{figure}

\begin{figure}
\epsscale{0.85}
\plotone{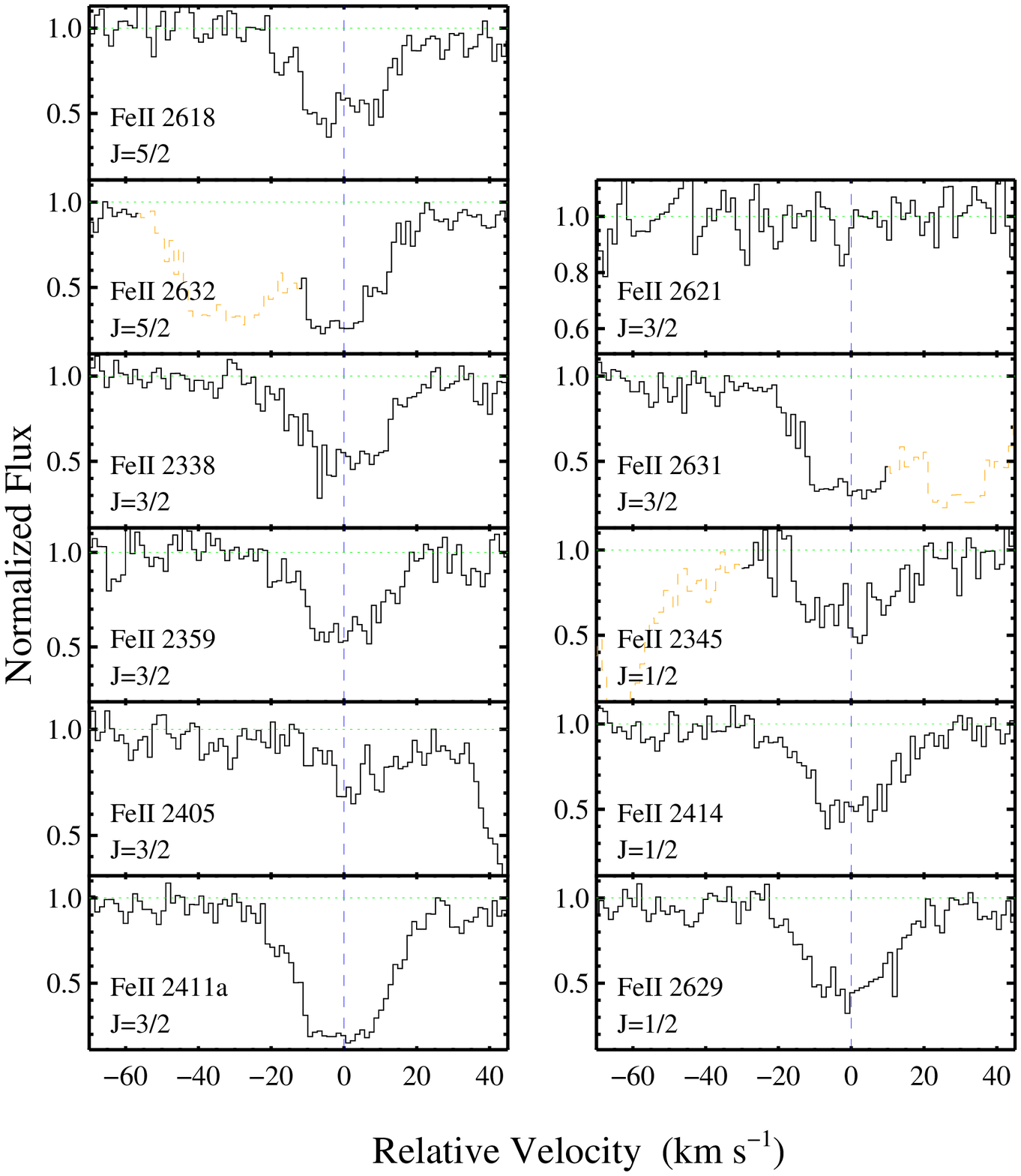}
\end{figure}
 
\begin{figure}
\epsscale{0.85}
\plotone{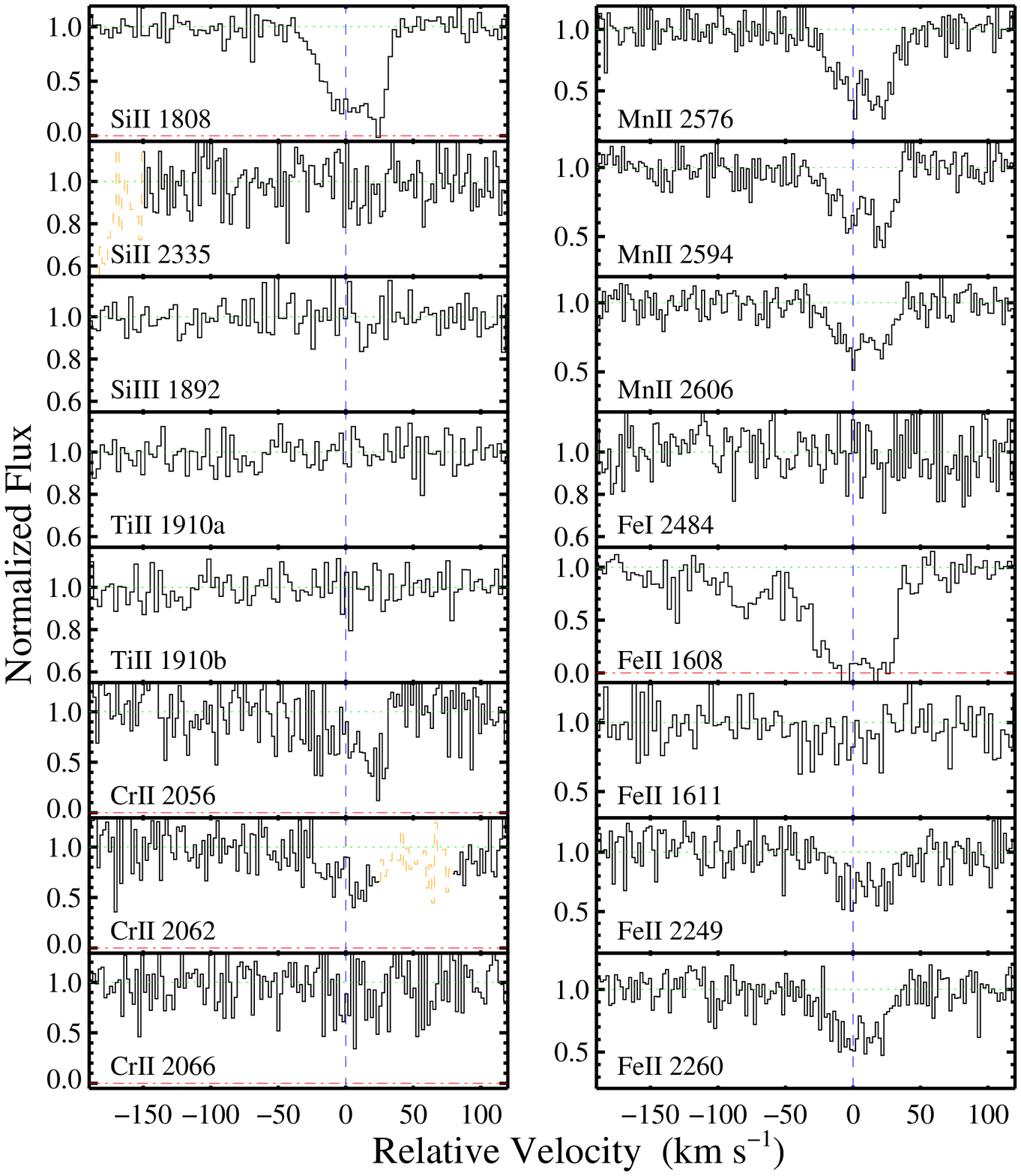}
\end{figure}

\begin{figure}
\epsscale{0.85}
\plotone{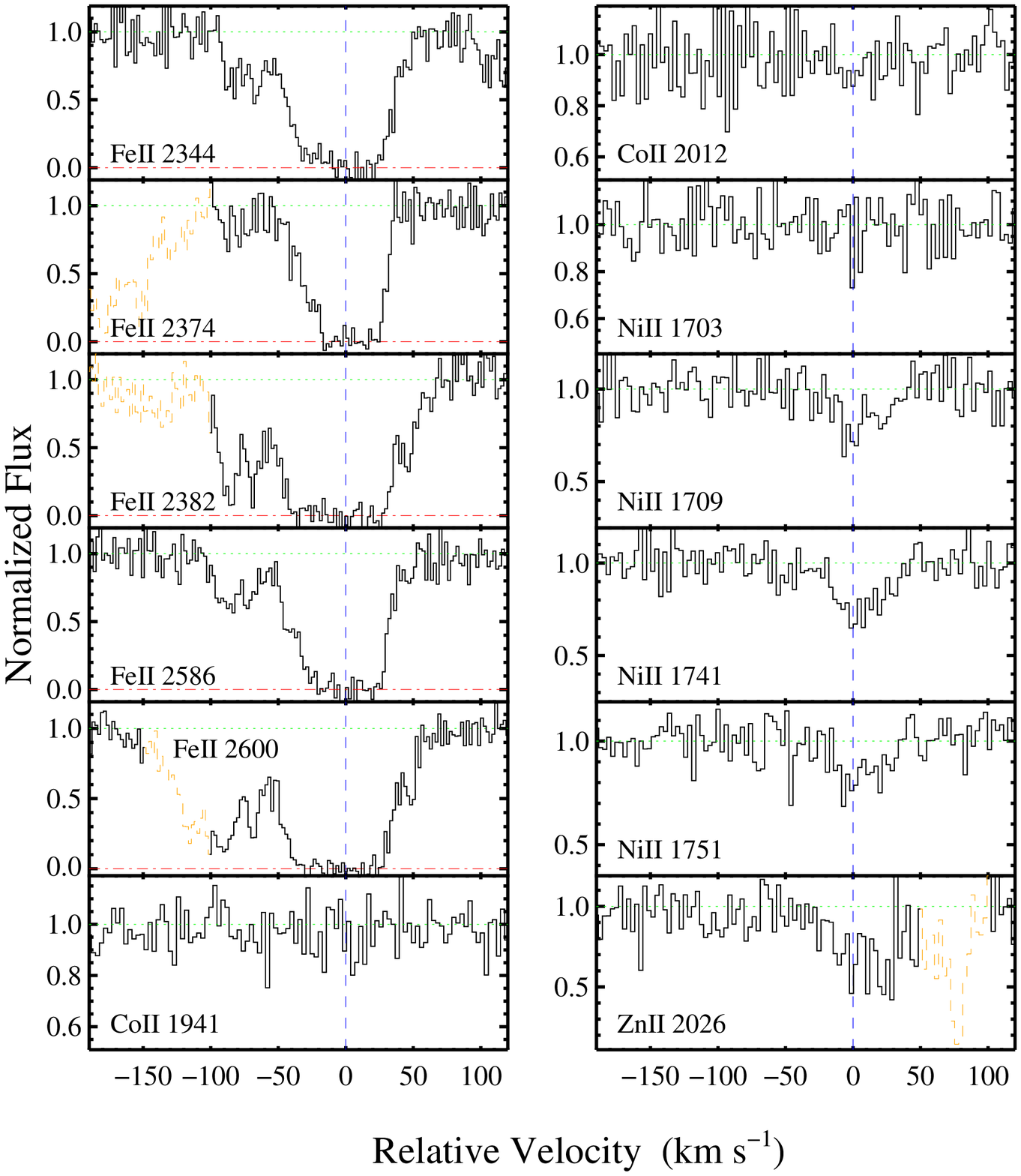}
\end{figure}

\begin{figure}
\epsscale{0.85}
\plotone{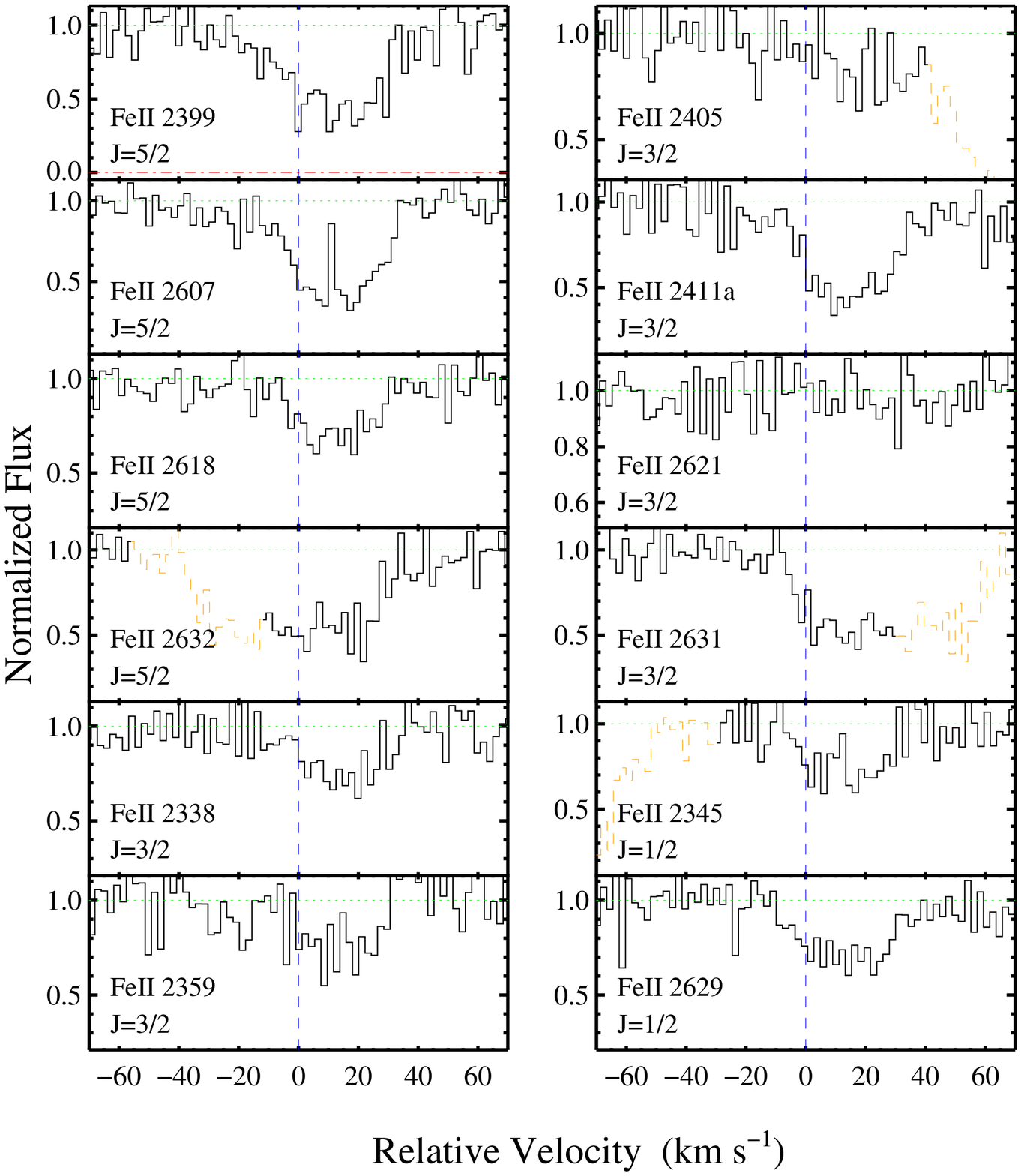}
\end{figure}
 
\end{document}